\documentclass[aps,prb,twocolumn]{revtex4}

\usepackage{amsmath}
\usepackage{amsthm}
\usepackage{amssymb}
\usepackage{graphicx}
\usepackage{tabularx}
\usepackage[squaren]{SIunits}

\let\a=\alpha    
   
\let\l=\lambda

\def\nn{\nonumber}

\def\be{\begin{equation}}
\def\ee{\end{equation}}
\def\bea{\begin{eqnarray}}
\def\eea{\end{eqnarray}}
\def\ba{\begin{array}}
\def\ea{\end{array}}

\newcommand{\bk}{{\bf k}}

\newcommand{\bq}{{\bf q}}                                                                                                                                            

\newcommand{\vk}{{\bf {k}}}

\newcommand{\vko}{{\bf k_1}}
\newcommand{\vkt}{{\bf {k_2}}}

\newcommand{\bg}{{\vec{ g}}}

\newcommand{\intk}{\int \frac{d^2\bk}{(2\pi)^2}}

\begin{document}

\title{Kinetic theory of Coulomb drag in two monolayers of graphene: \\ from the Dirac point to the Fermi liquid regime}

\author{Jonathan Lux}
\author{Lars Fritz}
\affiliation{Institut f\"ur Theoretische Physik, Universit\"at zu K\"oln, Z\"ulpicher Stra\ss e 77, 50937 K\"oln, Germany}

\date{\today}

\begin{abstract}
We theoretically investigate Coulomb drag in a system of two parallel monolayers of graphene. Using a Boltzmann equation approach we study a variety of limits ranging from the non-degenerate interaction dominated limit close to charge neutrality all the way to the Fermi liquid regime. In the non-degenerate limit we find that the presence of the passive layer can largely influence the conductivity of the active layer despite the absence of drag. This induces a non-trivial temperature behavior of the single layer conductivity and furthermore suggests a promising strategy towards increasing the role of inelastic scattering in future experiments. For small but finite chemical potential we find that the drag resistivity varies substantially as a function of the ratio of inelastic and elastic scattering. Furthermore, we explicitly show that the clean system has a well-defined drag resistivity even though the individual conductivities diverge. We find that an extrapolation from finite chemical potential to zero chemical potential and to the clean system is delicate and the order of limits matters. While the drag resistivity $\rho_d$ extrapolates to zero upon taking the limit $ \lim_{\overline{\alpha}\to \infty} \lim_{\mu_a=\mu_p \to 0} \rho_d=0$ it has a finite value in the opposite order of limits $\lim_{\mu_a=\mu_p \to 0} \lim_{\overline{\alpha}\to \infty} \rho_d=-\frac{1}{\sigma_0}$ ($\mu_a$ and $\mu_p$ are chemical potentials of the active and passive layer). The limiting value in the latter case is set by the interaction dominated single layer conductivity $\sigma_0$ of clean graphene and in that sense is a universal number. In the Fermi liquid regime we analyze drag as a function of temperature $T$ and the distance $d$ between the layers and compare our results to existing theoretical and experimental results. In addition to the conventional $1/d^4$-dependence with an associated $T^2$-behavior we find there is another regime of $1/d^5$-dependence where drag varies in linear-in-$T$ fashion. The relevant parameter separating these two regimes is given by $\overline{d}=T d/v_F$ ($v_F$ is the Fermi velocity), where $\overline{d} \ll1$ corresponds to $T^2$-behavior, while $\overline{d}\gg1$ corresponds to $T$-behavior. We speculate that the broad crossover between these two regimes was observed in recent experiments on graphene as well as old experiments on conventional two dimensional electron gases. We close with a discussion of the role of screening and the determination of the drag resistivity as function of the charge carrier densities in the two layers under very general circumstances covering the whole crossover from the non-degenerate to the degenerate limit in both layers independently.
\end{abstract}


\maketitle


\section{Introduction}

Graphene, a two dimensional system of carbon atoms arranged on a hexagonal lattice with an emerging Dirac type low-energy electronic dispersion continues to attract considerable interest on the theoretical and experimental front~\cite{RMPNeto}. One remarkable feature in experiments is that so far they have revealed only very limited information about interactions. The most prominent manifestations of interaction effects are the observation of the fractional quantum Hall effect~\cite{Andrei,Kim} as well as the logarithmic scaling of the Fermi velocity of the Dirac particles which was recently seen in quantum oscillation measurements on ultra-clean suspended samples~\cite{Guinea,Geim2011}. However, with ever increasing sample quality one expects to eventually be able to reach the hydrodynamic collision-dominated regime~\cite{Sheehy2007,Fritz2008,FosterAleiner,Kashuba2008} allowing to observe non-trivial many-body physics such as a collective cyclotron resonance~\cite{Mueller2008,Fritz22008} or an anomalously low viscosity~\cite{Fritz2009}. Also, a quantum-critical version of the Kondo effect possibly comes within reach~\cite{Fritz2011a}.  
A very direct manifestation of Coulomb interactions is provided by Coulomb drag experiments, the effect of electrons moving in one plane dragging along electrons in a plane parallel to the one in which the current is driven. This effect has a long history in the context of two dimensional electron gases~\cite{pogrebinskii1977,price1983,zheng1993,jauho1993,kamenev1995,flensberg1995,badalyan2007,asgari2008,rojo1999,gramila1991,sivan1992}. In graphene this problem has previously been studied in experiment~\cite{kim2011, kim2012,geim2012} and in a recent series of theoretical works~\cite{tse2007,narozhny2007,katsnelson2011,peres2011,hwang2011,narozhny2011,polini2012,schuett2012,levitov2012}. Here we report on theoretical results in the framework of a Boltzmann approach. Our approach goes beyond former approaches in that we allow for varying single layer properties as a function of all parameters. While this is not vital in the description of Coulomb drag in the Fermi liquid regime, $|\mu/T|\gg 1$, this becomes crucial in the non-degenerate limit, $|\mu/T|\ll1$, where interaction effects can dominate the single layer properties and an interesting interplay between elastic and inelastic scattering can be observed. Experimentally, there are indications that this regime should be within reach in experiments using samples prepared on hexagonal boron nitrid substrates where due to the atomically smooth surface that is relatively free of dangling bonds and charge traps high purity can be achieved and the puddle regime can be suppressed to very low densities~\cite{kim2010}. Our approach, like other theoretical approaches to date, is not valid in this strongly inhomogeneous regime. Throughout the paper we keep our results as general as possible, meaning we try to keep the number of Dirac cones $N$ in final expressions, if possible. This implies that our results should equally apply to three dimensional topological insulators, whose surfaces are characterized by an odd number of Dirac cones (in the case of weak topological insulators there is an even number of Dirac cones). A possible drag setup in such a system is even more straightforward and very natural in the sense that slab systems with a finite size gap in z-direction host a natural setting in which our results apply. However, we stress that the localization physics in these theories is different due to the helical nature of the surface Dirac fermions.

\subsection{General properties of drag}

\begin{figure}[h]
 \includegraphics[width=0.4\textwidth]{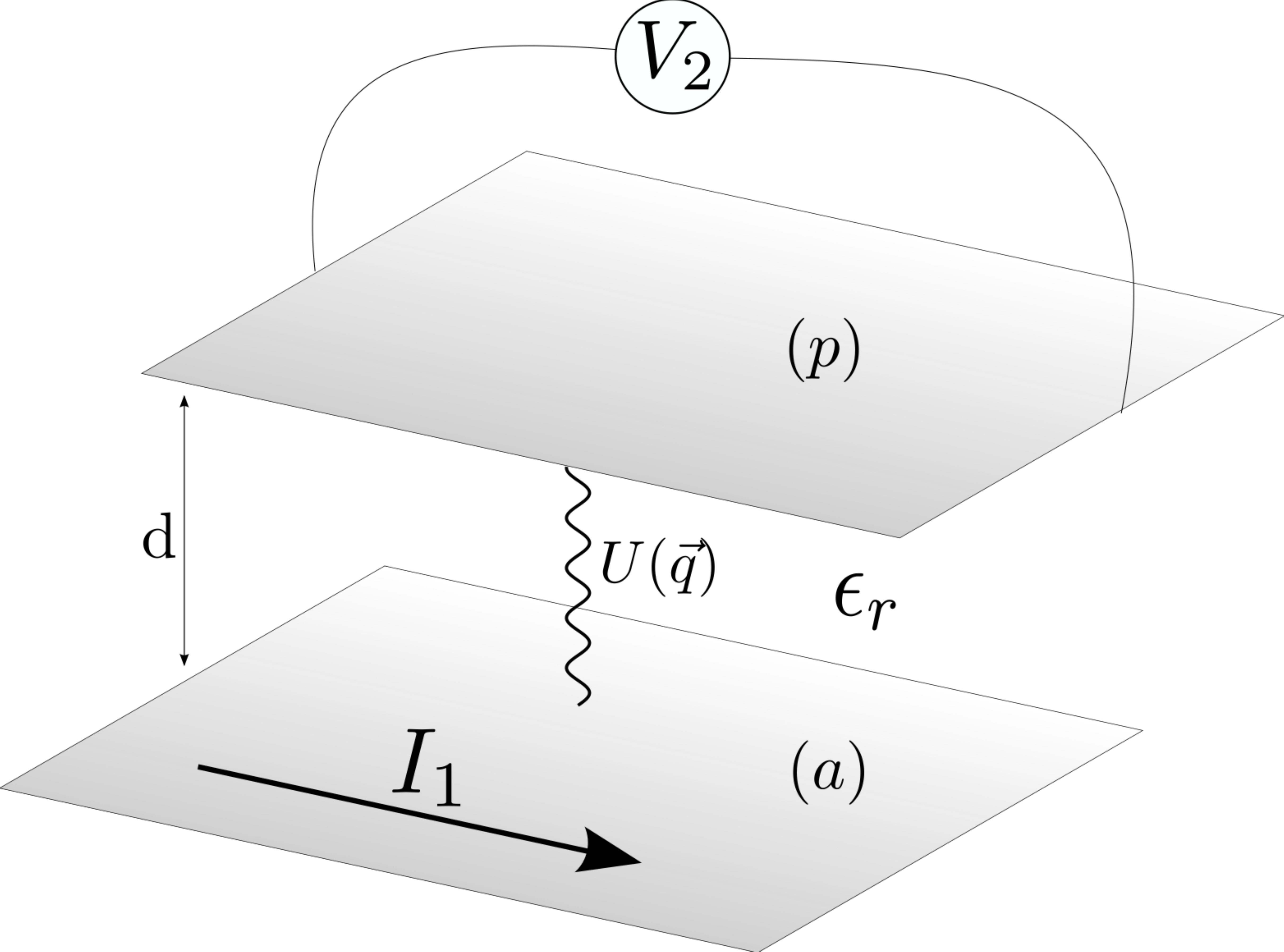}
\caption{Schematic setup of a drag experiment. In the active layer ($a$) a current $I_1$ is driven. In the passive layer (p) a voltage $V_2$ is induced such that overall there is no current flow in the passive layer. The drag resistance is defined as $R_2=-V_2/I_1$.}\label{Fig:fig1}
\end{figure}

In the experimental setup, Fig.~\ref{Fig:fig1}, two monolayers of graphene are separated by a distance $d$. We assume that in-between the monolayers there is an insulating region filled with a dielectric with a dielectric constant $\epsilon_r$. Throughout the paper the dielectric constant $\epsilon_r$ is not a function of the vertical coordinate. This situation has been studied elsewhere~\cite{polini2012}. We assume that the two layers can be individually gated such that the carrier concentration in both layers can be adjusted independently. Furthermore, we divide the two layers into active and passive layer, where active layer refers to the fact that within this layer a current is driven, while the passive layer will not carry current. In a standard experiment a current $I_1$ is driven through the active layer. 
If no current is allowed to flow in the passive layer this induces a voltage difference $V_2$, allowing to define a drag resistance $R_2=-V_2/I_1$.

We consider the response tensor which has a structure similar to the one in the Hall effect. We apply the electric field ${\bf{E}}_a$ only in the active layer $a$ but induce a current in the active layer $a$ called ${\bf{j}}_a$ as well as ${\bf{j}}_p$ in the passive layer $p$. Consequently, there are layer-diagonal and off-diagonal conductivities involved:
\begin{eqnarray}
\left ( \begin{array} c {\bf{j}}_a \\  {\bf{j}}_p \end{array} \right) = \left (  \begin{array} {cc}  \sigma_a & \sigma_d \\  \sigma_d & \sigma_{p}   \end{array}  \right) \cdot \left ( \begin{array} c {\bf{E}}_a \\  {\bf{0}} \end{array} \right) \;,
\end{eqnarray}
which includes the individual conductivities $\sigma_a$ of the active and that of the passive layer, $\sigma_p$, while the drag conductivity is denoted $\sigma_d$. In the concrete experiment, however, the boundary conditions are different and the passive layer does not carry current. Demanding ${\bf{j}}_p={\bf{0}}$ requires a field counteracting the flow in the passive layer which is given by ${\bf{E}}_p=-\frac{\sigma_d}{\sigma_p}{\bf{E}}_a$. This implies that the drag resistivity (or transresisitivity) is given by
\begin{eqnarray}\label{eq:drag}
\rho_d=\frac{|{\bf{E}}_p|^2}{{\bf{j}}_a \cdot {\bf{E}}_p}=\frac{-\sigma_d}{\sigma_a \sigma_p-\sigma_d^2}\;.
\end{eqnarray}

It is important to realize that like in the case of thermal transport $\rho_d$ can be finite even if the individual conductivities $\sigma_a$, $\sigma_p$, and $\sigma_d$ diverge, which we show explicitly. This is an effect of the boundary condition of vanishing charge flow in the passive layer analogous to a finite thermal conductivity in thermal transport in Fermi liquids.

\subsection{Summary of results}

Graphene bilayers turn out to provide an exceptionally versatile arena in which one can theoretically as well as experimentally vary a large number of parameters: (i) the temperature T, (ii) the chemical potentials of the individual layers $\mu_{a}$ and $\mu_{p}$, (iii) the interlayer spacing $d$, as well as (iv) disorder and (v) interaction strength via the dielectric environment. Within this work we do not attempt to exhaust all the possibilities offered by the above parameters but instead concentrate on the most interesting situations. Our main results concern among others the interplay of interactions and disorder in the limit of small chemical potentials, the so-called non-degenerate limit where $\left | \mu_{a} \right| ,\left|\mu_{p} \right|\ll T$ in both layers individually. We furthermore describe the crossover from this interesting limit to the more conventional Fermi liquid regime $\left |\mu_{a} \right|, \left|\mu_{p} \right|\gg T$. 

On a different note we study the dependence on distance $d$ in some detail in both the non-degenerate and the degenerate limit. In experiments distances $d$ can be realized for which the regime $d \ll v_F/T$ can be reached in absence of leakage currents for reasonable temperatures. This is an interesting limit especially in the non-degenerate case since the typical momentum of electrons involved in scattering events is on the order $T/v_F$ and consequently the interlayer Coulomb interaction can be considered essentially undamped, see Eq.~\eqref{eq:yukawa}, meaning we have the maximal effect of the inter-layer interaction strength and inelastic scattering can effectively be increased. However, we also study the opposite limit, $d \gg v_F/T$, which is particularly interesting in the Fermi liquid regime revealing a formerly not discussed regime.

Our main findings are as follows: (i) For zero doping in both layers, {\it i.e.}, chemical potential $\mu_{a,p}=0$, the passive layer remains in equilibrium and consequently the drag resistance is zero. This happens by virtue of the current carrying state being of zero total momentum which is enforced by particle-hole symmetry. Despite the passive layer remaining in equilibrium we find that its presence still has a large effect on the active layer. In the limit $T d/v_F \ll 1$, the conductivity of the active layer is reduced by roughly a factor of two. This comes about due to scattering of the electrons and holes in the active layer from plasmons in the passive layer which becomes increasingly pronounced upon decreasing temperature. The effect is unusually large since in a normal Fermi liquid the influence of the passive layer on the transport properties of the active layer is negligibly small. Overall, we find that the active layer shows a non-trivial temperature dependence of the conductivity. This provides a potential route towards increasing inelastic scattering in the active layer in a setup where the active layer is sandwiched between a (possibly large) number of passive layers which act as a reservoir for inelastic scattering. 
 (ii) In the non-degenerate limit $|\mu_a|,|\mu_p|\neq 0 \ll T$ we find an interesting crossover from disorder dominated drag to interaction dominated drag. In the limit of zero disorder, in which all individual conductivities diverge, the drag resistivity remains finite, which we show explicitly. Furthermore, we find that the approach to the clean limit combined with the approach of the Dirac point is subtle and the order of limits matters: this means that taking disorder to zero and subsequently the chemical potential to zero or the other way around yields differing results. In the former case one obtains a finite drag resistivity even at the Dirac point which is given by the inverse single layer conductivity of the clean system. In the latter order of limits one obtains zero. (iii) In the Fermi liquid regime we study the dependence of the transresistivity on the distance $d$ between the two layers. We find that there is an interesting crossover in the behavior of the drag resistivity as a function of the distance $d$ which has previously not been discussed in the context of Fermi liquids~\cite{kamenev1995,flensberg1995}, namely a change from $1/d^4$-behavior to $1/d^5$-behavior. This effect is accompanied by a crossover in the temperature dependence which goes from $T^2$ to $T$, which could be relevant for the understanding of recent experiments in the Fermi liquid regime as well as older experiments on conventional two-dimensional electron gases. As a byproduct we derive the standard formula for Coulomb drag in the Fermi liquid regime~\cite{kamenev1995,flensberg1995} from a very simple one-mode approach to the Boltzmann equation, which to the best of our knowledge has not been discussed before. (iv) We describe the full crossover from the non-degenerate limit into the Fermi liquid regime $|\mu_a|,|\mu_p|\gg T$ in which the single layer conductivities are disorder dominated. We find that the effect of screening in this limit brings us towards orders of magnitude of the drag resistivity which are very compatible with experimental results.

Technically we use the kinetic approach, which requires the full numerical solution of coupled Boltzmann equations for the distribution functions of the electrons and holes in the individual layers, thus for four coupled Boltzmann equations. This is a straightforward but non-standard application of the variational principle~\cite{Ziman} and consequently explained in some detail. The present work goes beyond former theoretical works in mainly three aspects: (a) We do not use a relaxation time approximation but instead solve the Boltzmann equation numerically within a two-mode approach (this is logarithmically exact in the strong coupling limit). (b) The description of the interplay of interactions and disorder especially in the non-degenerate limit is facilitated by the two-mode description, which is the minimal number of modes required for a faithful account. (c) We do not take the individual layer conductivities as input parameters but instead calculate them for every set of parameters which leads to qualitative and quantitative changes in the non-degenerate limit.

\subsection{Organization of the paper}

The organization of the paper is such that we start with a discussion of the setup in Sec.~\ref{sec:system} , which includes a discussion of the model Hamiltonian (Sec.~\ref{sec:model}). In Sec.~\ref{sec:crtss} we first discuss the sources of current relaxation in Sec.~\ref{sec:cr}, then the associated time scales in Sec.~\ref{sec:ts}, as well as the effect of screening in Sec.~\ref{sec:sc}. The generic framework of the Boltzmann equation and the matrix formalism used follows in Sec.~\ref{sec:kinetic}. We first introduce the coupled kinetic equations necessary to describe drag in Sec.~\ref{sec:cke}. Here we also explain how the effect of drag manifests itself in the structure of the coupled equations in linear response. We then move towards the variational ansatz in Sec.~\ref{sec:va} and shortly review the variational principle. We also discuss the minimal number of modes required for a faithful description within our problem. In a last step, Sec.~\ref{sec:mf}, we introduce a generic matrix formalism derived from the variational principle which enables us to calculate drag from an inversion of a matrix. Readers not interested in technical details may skip Sec.~\ref{sec:kinetic} and directly move to the results: we start with a discussion of the non-degenerate limit in Sec.~\ref{sec:nd}. In a first step we discuss the case of both layers at the Dirac point in Sec.~\ref{sec:bd}. The results lead us to propose an experimental setup, in which the effect of inelastic scattering can be increased considerably in Sec.~\ref{sec:increased}. We then move to finite chemical potential, but still $T \gg \mu$, in Sec.~\ref{sec:fcp}. There we discuss that the transresistivity can be finite even in a clean limit and show that drag largely depends upon the ratio of elastic to inelastic scattering. We furthermore discuss that the extrapolation to zero density and zero impurity density is delicate and the order of limits matters. In a next step we analyze the Fermi liquid regime in Sec.~\ref{sec:fl} which is what has been analyzed predominantly in other works. We first give an alternative derivation of the standard results of drag in Fermi liquid theory~\cite{kamenev1995,flensberg1995,narozhny2011,polini2012} obtained in the Kubo approach in terms of coupled Boltzmann equations. Then we discuss a number of different situations and study the behavior of drag with distance $d$ in great detail. Importantly, we find a regime of temperature and distance dependence which has been overlooked in previous works and could be important for the proper interpretation of recent experiments~\cite{kim2011,kim2012} as well as old experiments~\cite{gramila1991}. We continue our discussion in Sec.~\ref{sec:crossovers} where we derive full crossover functions for realistic setups covering the full range of chemical potentials with a particular emphasis on screening effects. We close with the conclusions in Sec.~\ref{sec:conclusion}. We have relegated a technical discussion of the full scattering kernel as well as the matrix elements of the scattering matrices to the Appendix.

\section{The model, time scales, and screening}\label{sec:system}
\subsection{The model} \label{sec:model}

The model Hamiltonian consists of two copies of the free graphene Hamiltonian for the active and passive layer, respectively, and interactions within and in-between layers. It reads
\begin{eqnarray}
H = \sum_{i=a,p} \left( H_0^i + H_{\rm{int}}^i + H_{\rm{dis}}^i  \right)+H^{ap}_{\rm{int}},
\end{eqnarray}
where $a$ denotes the active layer and $p$ the passive. $H_0^{a/p}$ denotes the free Hamiltonian in both layers, $H_{\rm{int}}^{a/p}$ the interaction within each layer, while $H^{ap}_{\rm{int}}$ describes the interaction between layers. Disorder is implemented within each layer via $H_{\rm{int}}^i$. The non-interacting Hamiltonian reads
\begin{eqnarray}
H_0^{i} =- \sum_{f=1}^N\int d^2 \mathbf{x} \left[  \Psi_f^{i\dagger} \left( iv_F \vec{\sigma}\cdot \vec{\nabla}
-\mu_i  \right) \Psi^{i\phantom{\dagger}}_f \right]\;,
\end{eqnarray}
with the Fermi velocity $v_F$, $f=1,...,N$ counting the flavors, and $\mu_i$ being the chemical potential of the individual layers. In the case of graphene we have $N=4$ due to valley and spin degeneracy, while for a topological insulator we would rather expect $N=1$ (or more generally an odd integer). The spinor representation of the wave-function has the following Fourier decomposition
\begin{equation}
\Psi^{i\phantom{\dagger}}_f(\mathbf{x},t)=\int \frac{d^{2}k}{(2\pi )^{2}}\left(
\begin{array}{c}
c^i_{1f}(\mathbf{k},t) \\
c^i_{2f}(\mathbf{k},t)
\end{array}
\right) e^{i\mathbf{k}\cdot \mathbf{x}},
\end{equation}
where the operators $c^i_{1/2f}$ are the electron annihilation operators on the two different sublattices for flavor index $f$ and in layer $i=a,p$. We note that in topological insulators the spinorial components do not refer to the sublattice but rather to the spin degree of freedom accounting for their helical nature.
The formulation of transport is simplest in a basis which
diagonalizes the Hamiltonian $H_{0}^i$. This is accomplished by a unitary transformation from the Fourier mode operators $
(c^i_{1f},c^i_{2f})$ to the basis of electrons and holes $(\gamma _{+a},\gamma _{-a})$:
\begin{eqnarray}
c^i_{1f}(k) &=&\frac{1}{\sqrt{2}}(\gamma ^i_{+f}(\mathbf{k})+\gamma^i_{-f}(
\mathbf{k})),  \notag  \label{eq:unitary} \\
c^i_{2f}(k) &=&\frac{K}{\sqrt{2}k}(\gamma^i_{+f}(\mathbf{k})-\gamma^i_{-f}(
\mathbf{k})).
\end{eqnarray}
Above, we introduced the complex number $K$ by the relation
\begin{equation}
K\equiv k_{x}+ik_{y},~~~~\mbox{where}~~~~~\mathbf{k}\equiv (k_{x},k_{y}),
\end{equation}
and $k=|\mathbf{k}|=|K|$. Expressing the Hamiltonian $H_{0}^i$ in terms of $
\gamma^i_{\pm f}$, we obtain
\begin{equation}
H^i_{0}=\sum_{\lambda=\pm}\sum_ {f=1}^N\int \frac{d^{2}k}{(2\pi )^{2}}\lambda v_{F}k\, \gamma
_{\lambda f}^{i\dagger }(\mathbf{k})\gamma^{i\phantom{\dagger}}_{\lambda f}(\mathbf{k})\;.
\end{equation}

The distribution functions of electrons and holes ($\pm$) in the layers $i=a/p$ read
\begin{equation}
f^i_{\lambda }(\mathbf{k},t)=\left\langle \gamma_{\lambda f}^{i\dagger }(%
\mathbf{k},t)\gamma_{\lambda f}^{i\phantom{\dagger}}(\mathbf{k},t)\right\rangle .  \label{defg}
\end{equation}%
There is no sum over $f$ on the right hand side, and we assume the distribution
functions to be the same for all valleys and spins, which is why we drop the index $f$ from now on. In equilibrium, {\it i.e.}, in the
absence of external perturbations, the distribution functions are Fermi-Dirac functions
\begin{eqnarray}
f^i_{\lambda}(\mathbf{k},t) &=&f^{0}_{\lambda}(v_{F}k)= \frac{1}{e^{\frac{\lambda v_F k-\mu}{T}}+1}\;.
\end{eqnarray}%
The current can be expressed in terms of the electron- and hole-operators and decomposes into
\begin{equation}
\mathbf{J}=\mathbf{J}_{I}+\mathbf{J}_{II}
\end{equation}%
with
\begin{equation}
\mathbf{J}_{I}=ev_{F}\sum_{\lambda a}\int \frac{d^{2}k}{(2\pi )^{2}}\frac{%
\lambda \mathbf{k}}{k}\gamma _{\lambda a}^{\dagger }(\mathbf{k})\gamma
_{\lambda a}(\mathbf{k})\,,  \label{defj1}
\end{equation}%
and
\begin{eqnarray}
\mathbf{J}_{II}&=&-iev_{F}\int \frac{d^{2}k}{(2\pi )^{2}}\frac{(\hat{\mathbf{z}%
}\times \mathbf{k})}{k} \nonumber \\ &\times& \left[ \gamma _{+a}^{\dagger }(\mathbf{k})\gamma
_{-a}(\mathbf{k})-\gamma _{-a}^{\dagger }(\mathbf{k})\gamma _{+a}(\mathbf{k}%
)\right]\,, \label{defj2}
\end{eqnarray}%
where $\hat{\mathbf{z}}$ is a unit vector orthogonal to the $x,y$ plane.
$\mathbf{J}_{I}$ measures the current carried by motion of the
quasiparticles and quasiholes---notice the $\lambda $ prefactor, indicating
that these excitations have opposite charges. The operator $\mathbf{J}_{II}$
creates a quasiparticle-quasihole pair (it corresponds to the so-called Zitterbewegung, see Ref.~\cite{RMPNeto}) and is the part which determines the optical conductivity. For the purpose of this paper we can neglect its influence on transport properties, since we are interested in d.c. transport properties. In the framework of the Kubo formula, which fully accounts for the off-diagonal parts, it was shown that this leads to numerically identical results~\cite{schuett}.

In a particle-hole symmetric situation a current carrying state with
holes and electrons moving in opposite directions has a
vanishing total momentum, and the current can decay by creation or annihilation of particle hole pairs, without violation of momentum conservation. This is the physical reason why at the
particle hole symmetric point, {\it i.e.}, at vanishing  deviation of the chemical
potential from the Dirac point, the d.c. conductivity is finite even in the
absence of momentum relaxing impurities. However, as we will see below, at finite
deviation from particle hole symmetry a driving electric field always excites
the system into a state with finite momentum which cannot decay. This 
entails an infinite d.c. conductivity (even though drag can be finite), and consequently impurities have to be taken into account.

\subsection{Sources of current relaxation, time-scales, and screening}\label{sec:crtss}

In the following we discuss three important ingredients for our subsequent discussions, which are disorder effects, interaction effects, as well as screening properties in two dimensional Dirac systems.

\subsubsection{Sources of current relaxation} \label{sec:cr}

Within this work we study the interplay of three different sources of current relaxation, which are intralayer Coulomb interaction, interlayer Coulomb interaction, and disorder.

The $1/r$ intralayer Coulomb interaction assumes the form 
\begin{eqnarray}
&& H^i_{\rm{int}} = \sum_{f,f'=1}^N \sum_{\lambda_1 \lambda_2 \lambda_3 \lambda_4} \int \frac{d^2 k_1 }{
(2 \pi )^2} \frac{d^2 k_2 }{(2 \pi )^2} \frac{d^2 q }{(2 \pi )^2}  \\
&&\times T_{\lambda_1 \lambda_2 \lambda_3 \lambda_4} (
\mathbf{k}_1 , \mathbf{k}_2 , \mathbf{q} ) \gamma_{\lambda_4 f'}^{i\dagger} (
\mathbf{k}_1+\mathbf{q} ) \gamma_{\lambda_3 f}^{i\dagger} ( \mathbf{k}_2-
\mathbf{q} ) \nonumber \\ &&~~~~~~~~~~~~~~~~~\times \gamma^i_{\lambda_2 f} ( \mathbf{k}_2 ) \gamma^i_{\lambda_1 f'} (
\mathbf{k}_1 )\,.\nonumber
\end{eqnarray}
Here the scattering matrix element reads
\begin{eqnarray}
&&T_{\lambda_1 \lambda_2 \lambda_3 \lambda_4} (\mathbf{k}_1 , \mathbf{k}_2 ,
\mathbf{q}) = \frac{V({\bf q},\omega_{\mathbf{k}_1 ,\mathbf{q}})}{8} \times  \\ &\times& \left[ 1 +
\lambda_1 \lambda_4 \frac{(K_1^{\ast} + Q^{\ast}) K_1}{|\mathbf{k}_1 +
\mathbf{q}| k_1} \right] \left[1 + \lambda_2 \lambda_3 \frac{(K_2^{\ast} -
Q^{\ast}) K_2 }{|\mathbf{k}_2 - \mathbf{q}| k_2} \right] , \nonumber  \label{deft}
\end{eqnarray}
where $\omega_{\mathbf{k}_1,\mathbf{q}}= v_F(\lambda_4|\mathbf{k}_1 +
\mathbf{q}|-\lambda_1|\mathbf{k}_1|)$, and
\begin{eqnarray}
V({\bf q},\omega)=\frac{2\pi e^2}{\epsilon_r |{\bf q}|}
\end{eqnarray}
is the dynamically screened Coulomb interaction. In this expression $\epsilon_r$ is the dielectric constant of the adjacent media. Note that we have neglected the scattering between valleys since it connects points in the Brillouin zone which involve large momentum transfers and consequently are strongly suppressed. The two layers are at a vertical distance $d$ (in z-direction) and consequently the unscreened interlayer Coulomb interaction reads
 \begin{eqnarray}
 U({\bf{r}})\propto \frac{1}{\sqrt{{\bf{r}}^2+d^2}}
 \end{eqnarray}
which after Fourier transform assumes the form
\begin{eqnarray}\label{eq:yukawa}
U({\bf q},\omega)=\frac{2\pi e^2}{\epsilon_r |{\bf q}|}e^{-qd}\;.
\end{eqnarray}
The Hamiltonian $H^{ap}_{\rm{int}}$ which connects the two layers assumes the following form in the basis of electrons and holes 
\begin{eqnarray}
&& H^{ap}_{\rm{int}} = \sum_{f,f'=1}^N \sum_{\lambda_1 \lambda_2 \lambda_3 \lambda_4} \int \frac{d^2 k_1 }{
(2 \pi )^2} \frac{d^2 k_2 }{(2 \pi )^2} \frac{d^2 q }{(2 \pi )^2}  \\
&&\times \tilde{T}_{\lambda_1 \lambda_2 \lambda_3 \lambda_4} (
\mathbf{k}_1 , \mathbf{k}_2 , \mathbf{q} ) \gamma_{\lambda_4 f'}^{a\dagger} (
\mathbf{k}_1+\mathbf{q} ) \gamma_{\lambda_3 f}^{p\dagger} ( \mathbf{k}_2-
\mathbf{q} ) \nonumber \\ &&~~~~~~~~~~~~~~~~~\times \gamma^p_{\lambda_2 f} ( \mathbf{k}_2 ) \gamma^a_{\lambda_1 f'} (
\mathbf{k}_1 )\nonumber
\end{eqnarray}

with
\begin{eqnarray}\label{eq:tmatrix}
&&\tilde{T}_{\lambda_1 \lambda_2 \lambda_3 \lambda_4} (\mathbf{k}_1 , \mathbf{k}_2 ,
\mathbf{q}) = \frac{U({\bf q},\omega_{\mathbf{k}_1 ,\mathbf{q}})}{8} \times  \\ &\times& \left[ 1 +
\lambda_1 \lambda_4 \frac{(K_1^{\ast} + Q^{\ast}) K_1}{|\mathbf{k}_1 +
\mathbf{q}| k_1} \right] \left[1 + \lambda_2 \lambda_3 \frac{(K_2^{\ast} -
Q^{\ast}) K_2 }{|\mathbf{k}_2 - \mathbf{q}| k_2} \right] . \nonumber  \label{inter}
\end{eqnarray}
In order to discuss situations away from the Dirac point we have to include the effect of disorder, which is required in order to obtain finite individual layer conductivities. This is required since at finite chemical potential an electric field excites a finite momentum state, which can only be relaxed due to translational invariance breaking. We assume the following form of the disorder potential
\begin{eqnarray}
H_{\textrm{dis}}^i= \sum_f \int d {\mathbf{x}}V_{\textrm{dis}}({\mathbf{x}})\Psi_f^{i\dagger}({\mathbf{x}}) \Psi^{i\phantom{\dagger}}_f ({\mathbf{x}})\;,
\end{eqnarray}
with
\begin{eqnarray}
V_{\rm dis} ({\bf x})= \sum_{
i} \frac{Z e^2}{\varepsilon |{\bf x}-{\bf x}_i|}.
\end{eqnarray}
Here ${\bf x}_i$ denotes the random positions of charged impurities, assumed to be close to the graphene sheet, having a charge $Z e$ and average spatial density $\rho_{\rm imp}$.
The disorder Hamiltonian $H_{\rm{dis}}$ in terms of the $
\gamma^i_{\lambda f}$ reads
\begin{eqnarray}
H^i_{\textrm{dis}}&=&\sum_{i}\sum_{a=1}^N \sum_{\lambda_1 \lambda_2} \frac{d^2 k_1 }{(2 \pi )^2} \frac{d^2 k_2 }{(2 \pi )^2} U_{\lambda_1\lambda_2} (\vko,\vkt) \\ &&\times \exp[i{\bf x}_i\cdot(\vko-\vkt)] \gamma^{i \dagger}_{\lambda_1 f}(\vko) \gamma^{i \phantom{\dagger}}_{\lambda_2 f}(\vkt),\nn
\end{eqnarray}
where
\begin{eqnarray}\label{dispot}
U_{\lambda_1\lambda_2}(\vko,\vkt)=-\frac{2\pi Z e^2}{\epsilon_r |\vko-\vkt|}\, \frac{1}{2}\left[ 1 +
\lambda_1 \lambda_2 \frac{K_1^\ast K_2}{k_1 k_2} \right]\, ,
\end{eqnarray}
which corresponds to unscreened Coulomb scatterers.
Note that even though we compute specific results for Coulomb interacting particles and Coulomb impurities, the formalism easily generalizes to arbitrary isotropic two body interactions and disorder potentials coupling to the local charge density. In the following we assume that disorder only acts within one layer and remains unscreened even at finite chemical potential. This does not influence any conclusions drawn from our analysis and using scalar impurity potentials would yield identical results. 
 
\subsubsection{Time scales} \label{sec:ts}

The transport timescales within a layer have been discussed before~\cite{Mueller2008} and we repeat the major results here. For a clean system at the Dirac point we find that electron-electron interactions induce a finite inelastic scattering rate. Introducing the 'fine structure constant'
\begin{eqnarray}
\alpha=\frac{e^2}{\epsilon_r v_F}
\end{eqnarray}
which has a logarithmic scaling~\cite{Guinea} we find that close to zero doping it is on the order of
\begin{eqnarray}
\tau^{-1}_{\rm ee}\sim {\alpha^2}\frac{k_BT}{\hbar},
\end{eqnarray}
and thus essentially set by the temperature. This is a hallmark of the quantum criticality of the undoped graphene system~\cite{Sheehy2007,Kashuba2008,Fritz2008}. The full crossover from quantum critical to Fermi liquid is described by
\begin{eqnarray}\label{eq:inter}
\tau^{-1}_{\rm ee}\sim {\alpha^2}\frac{k_BT^2/\hbar}{ {\rm max}[k_B T,\mu]}\;,
\end{eqnarray}
where at larger doping, when the chemical potential $\mu$ exceeds $k_BT$, the inelastic scattering rate tends to the expected Fermi liquid form $\tau^{-1}_{\rm ee}\sim T^2/\mu$, if screening is taken into account. 
The elastic scattering rate due to static charged impurities is naturally proportional to the density of impurities and in general reads
\begin{eqnarray}\label{eq:dis}
\tau^{-1}_{\rm imp}\sim \frac{1}{\hbar}\frac{(Ze^2/\epsilon_r)^2\rho_{\rm imp}}{{\rm max}[k_B T,\mu]}.
\end{eqnarray}
We note that the inelastic scattering rate decreases with temperature, while the elastic scattering rate increases. The latter is due to the fact that low energy particles are more intensely scattered by Coulomb impurities. Again, it is worthwhile mentioning that the physics of electron-hole puddles is beyond this description and our results do not apply in the inhomogeneous regime.

\subsubsection{The effect of screening}\label{sec:sc}

We introduce the two independent polarization functions $\Pi_a$ and $\Pi_p$ for the active layer and the passive layer, respectively. The random phase approximation (RPA) in the basis of intra- and interlayer interactions leads to the following Dyson equation~\cite{kamenev1995}
\begin{widetext}
\begin{eqnarray}
\left( \begin{array} {cc} V_{aa} \left( {\bf{q}},\omega \right) & U_{ap} \left( {\bf{q}},\omega \right) \\ U_{ap} \left( {\bf{q}},\omega \right)& V_{pp}\left( {\bf{q}},\omega \right)  \end{array} \right)=\left( \begin{array} {cc} V & U   \\ U & V  \end{array} \right) - \left( \begin{array} {cc} V & U   \\ U & V  \end{array} \right)  \left( \begin{array} {cc} -\Pi_a\left( {\bf{q}},\omega \right) & 0  \\ 0 & -\Pi_p\left( {\bf{q}},\omega \right)  \end{array} \right)\left( \begin{array} {cc} V_{aa}\left( {\bf{q}},\omega \right) & U_{ap}\left( {\bf{q}},\omega \right)  \\ U_{ap}\left( {\bf{q}},\omega \right) & V_{pp}\left( {\bf{q}},\omega \right)  \end{array} \right)
\end{eqnarray}
which can be solved in an elementary way yielding
\begin{eqnarray}\label{eq:RPA}
\left( \begin{array} {cc} V_{aa} & U_{ap}  \\ U_{ap}& V_{pp}  \end{array} \right)=\frac{1}{\left(1+V \Pi_a\right)\left(1+V \Pi_p\right)-U^2 \Pi_a \Pi_p   }\left(
\begin{array}{cc}
V+\left(V^2-U^2\right) \Pi _p & U \\
 U&
   V+\left(V^2-U^2\right) \Pi _a \\
\end{array}
\right)\;.
\end{eqnarray}
\end{widetext}

{\it $|\mu|/T \ll 1$: The non-degenerate limit.}

\noindent A peculiarity of a theory of massless Dirac fermions at the charge neutrality point is the absence of standard Thomas-Fermi screening. This can be rationalized from the absence of density of states at the Fermi level. This means only a thermal density of states enters. The zero temperature polarization in the Matsubara formulation reads~\cite{RMPNeto,wunsch2006,hwang2007}
\begin{eqnarray}
\Pi_{a,p} ({\bf{q}},\omega_n)  =   \frac{N{\bf{q}}^2}{16\sqrt{v_F^2 {\bf{q}}^2+\omega_n^2}} \;,
\end{eqnarray}
From the static limit $\pi({\bf{q}},\omega_n=0)\propto |{\bf{q}}|$ the absence of screening immediately follows. Taking into account the thermal density of electrons the polarization reads
\begin{eqnarray}
\Pi_{a,p}({\bf{q}},\omega_n=0,T,\mu_{ap})&\approx & \frac{N {\rm{max}}[T,\mu]}{2\pi v_F^2}+ \frac{N|{\bf{q}}|}{16v_F}\nonumber \\ &=&\frac{N T}{2\pi v_F^2}+ \frac{N|{\bf{q}}|}{16v_F}\;,
\end{eqnarray}
where $T$ plays the role of the Thomas-Fermi screening momentum. Since typical momenta involved in the scattering process in this regime are on the order $T/v_F$ we conclude that the screening only makes a small contribution. This contribution is again controlled in $\alpha$ which is small and thus to leading order can consistently be neglected. It turns out that in the hydrodynamic regime screening must only be taken seriously if one wants to go beyond the two-mode approximation. However, then the dynamic part is only important to cut off the forward scattering singularity~\cite{Fritz2008,Kashuba2008}.

{\it $|\mu|/T \gg 1$: The Fermi liquid regime.}

\noindent In this limit only one of the two charger carriers matters and consequently one can carry out a simplified analysis~\cite{kamenev1995}. The real part of the retarded polarization in the static limit is given by the density of states at the Fermi level and consequently reads
\begin{eqnarray}\label{eq:screen}
\operatorname{Re} \Pi_{a,p} ({\bf{q}},\omega_n=0,T,\mu_{a,p}) \approx \frac{N\mu_{a,p}}{2 \pi v_F^2}\;.
\end{eqnarray}
This will account for the static screening in the Fermi liquid regime and provides the standard Thomas-Fermi expression for the screening vector $q_{{\rm{TF}}}$. We will use this approximate form in the discussion of the Fermi liquid regime where we use it in $U_{ap}$ in Eq.~\eqref{eq:coulombfermi}. 

In order to understand the Fermi liquid regime and its limiting behavior starting from the analytical formula Eq.~\eqref{eq:rhodragfl} we also need the imaginary part of the retarded polarization, which is given by
\begin{eqnarray}
\operatorname{Im} \Pi_{a,p} \approx \frac{N\mu_{a,p}}{2 \pi v_F^2} \frac{\omega}{v_F q} \theta \left (v_F q-|\omega|  \right)\;.
\end{eqnarray}

We find that the correct polarization functions matter for both the determination of the correct distance and temperature behavior of the drag resistivity as well as for the orders of magnitude in the drag resistivity when compared to experiment.

\section{The kinetic approach for the bilayer system}\label{sec:kinetic}

The Boltzmann equation approach has been used in the context of single-layer graphene in the collision-dominated hydrodynamic limit before~\cite{Fritz2008}.
We assume that the quasiparticle description remains valid in all regions of interest in our problem. The equation of motion for the quasiparticle distribution function schematically reads
\begin{eqnarray}
\partial_t f - {\bf{F}}_{\rm{ext}} \partial_{\bf{k}}f=-I_{\rm{coll}}
\end{eqnarray}
where $f$ is the distribution function, ${\bf{F}}_{\rm{ext}}$ denotes the external force, $\partial_t$ accounts for some temporal variation, and $I_{\rm{coll}}$ is the collision integral. 

In our case, the system under investigation has the generic form shown in Fig.~\ref{Fig:fig1} and consequently requires to extend the formalism of the single layer to also account for the presence of the passive layer and interactions between the two layers. This leads to a total of four coupled equations of motion which have to be solved simultaneously. Again, we can restrict our analysis to only include the diagonal parts of the distribution matrix. This is justified since we are only interested in d.c. properties. For optical properties this would not be justified.

\subsection{Coupled kinetic equations}\label{sec:cke}

The general structure of the coupled Boltzmann equations in the stationary limit, $\partial_t f=0$, assumes the form
\begin{eqnarray}\label{eq:boltz}
-e {\bf{E}}\nabla _{\mathbf{k}
} f^a_{+ }&=& -I_{\rm{C}}^{aa} - I_{\rm{C}}^{ap}-I_{\rm{dis}}^{aa} \nonumber \\ -e {\bf{E}}\nabla _{\mathbf{k}
}f^a_{- }&=&-I_{\rm{C}}^{aa} - I_{\rm{C}}^{ap}-I_{\rm{dis}}^{aa} \nonumber \\ 0&=& -I_{\rm{C}}^{pp} - I_{\rm{C}}^{pa}-I_{\rm{dis}}^{pp} \nonumber \\ 0&=& -I_{\rm{C}}^{pp} - I_{\rm{C}}^{pa}-I_{\rm{dis}}^{pp}\;.
\end{eqnarray}
The two uppermost lines account for the active layer in which both electrons and holes are subject to an applied electrical field. The lower two lines account for the passive layer, in which no field is applied requiring the left-hand side to be zero. There is a number of collision terms, where $I_{\rm{C}}^{aa}$ and $I_{\rm{C}}^{pp}$ account for the scattering due to Coulomb interaction within a layer ($a$ and $p$ stand for active and passive layer respectively), $I_{\rm{C}}^{ap}$ and $I_{\rm{C}}^{pa}$ account for inter-layer scattering, while $I_{\rm{dis}}^{aa}$ and $I_{\rm{dis}}^{pp}$ denote scattering due to disorder within the individual layers. The explicit form of the collision terms is presented in Appendix~\ref{App:collision} while the matrix elements of the scattering matrix are defined in Appendix~\ref{App:scatt}. Scattering between the active and the passive layer only includes processes which are of the density-density (large-$N$) type. One could faithfully describe this by an effective plasmonic mode for the passive layer coupling to the active layer thereby reducing the number of degrees of freedom~\cite{Fritz2011}. However, we choose to work in the basis described in Eq.~\eqref{eq:boltz}.
The effect of drag can easily be understood from the Boltzmann equation. In linear response the distribution functions in the active layer $f^a_{\pm}$ are driven out of equilibrium linearly in the applied field. Consequently, we have to plugging this ansatz into the lower two lines the term $I_{\rm{C}}^{ap}$. This implies that now the lower two lines also include a part which is linear in the applied field. This indirectly serves as a 'source term' for the distribution functions $f_{\pm}^p$ in the passive layer. In order to solve the lower two Boltzmann equations in linear response it follows that we now have to choose the deviation of $f_{\pm}^p$ from equilibrium to also be linear in the applied field in the active layer. In a Kubo formula approach this effect is captured by the standard Aslamazov and Larkin diagrams~\cite{varlamovlarkinbook}.

\subsection{Variational ansatz and choice of modes}\label{sec:va}

As discussed in Sec.~\ref{sec:cke}, the distribution function of the quasiparticles in both layers have to be expanded to linear order in the applied electrical field and consequently assume the schematic form
\begin{eqnarray}\label{eq:lr}
f^{a/p}_{\pm} &=&f^{0,a/p}_{\pm}(v_{F}k) \nonumber \\ &+&  \frac{ev_F}{T^2} \frac{{\bf{k}}}{|{\bf{k}}|}\cdot{\bf{E}}f^{0,a/p}_{\pm}\left (1-f^{0,a/p}_{\pm} \right) \chi^{a/p}_{\pm} \left ( \frac{v_F k}{T}\right) \nonumber \\
\end{eqnarray}
which provides the starting point of the subsequent discussion. The solution strategy is to choose an ansatz for the functions $ \chi^{a/p}_{\pm}$ which is related to the slow modes in the problem. In the non-degenerate limit the analysis requires only one mode to yield an asymptotically exact result~\cite{Fritz2008}. In the degenerate limit with $\mu/T\gg1$ the most important mode is the momentum mode. Both modes share the property that they can annihilate the divergence in the forward scattering amplitude of the Coulomb collision kernel, which is a peculiarity of electrons with linear dispersion in two dimension\cite{Fritz2008,Kashuba2008}. These modes thus constitute the leading contribution to current relaxation to leading logarithmic accuracy, which has been used before to describe the crossover of the single layer conductivity from the non-degenerate limit to the degenerate limit~\cite{Mueller2008}. For a in depth discussion of the logarithmic singularity in forward scattering we refer the reader to Ref.~\cite{Mueller2008}. The appropriate minimal ansatz for our purposes consequently is given by
\begin{eqnarray}\label{eq:ansatz}
\chi^{a/p}_{\pm}\left( \frac{v_F k}{T} \right)= \pm \chi^{a/p}_{0\pm}+\chi^{a/p}_{1\pm}\frac{v_F k}{T}
\end{eqnarray}
where $\chi_0$ is associated with particle-hole symmetry and the $\pm$ accounts for that while $\chi_1$ refers to the momentum conservation. The mode $\chi_0$ dominates transport in the non-degenerate limit. On the other hand $\chi_1$ dominates in the degenerate limit. The solution of the Boltzmann equation now is equivalent to determining the coefficients $\chi^{a/p}_{0\pm}$ and $\chi^{a/p}_{1\pm}$, which can be mapped to a matrix inversion problem. The general formalism is an application of the variational principle for coupled Boltzmann equations~\cite{Ziman} which is explained in great detail below.

\subsection{Matrix formalism for drag}\label{sec:mf}

Using the set of functions defined in Eq.~\eqref{eq:ansatz} the Boltzmann equation and its solution can be mapped to a matrix inversion, where the matrix acts in a combined space of layer indices, electrons, holes, and modes. This gives access to the expansion coefficients $\chi^{a/p}_{0,\pm}$ and $\chi^{a/p}_{1,\pm}$, which then allows to determine the individual and trans-conductivities. The major numerical effort within this approach goes into a faithful calculation of the matrix elements of the collision kernel. The space of functions, layers, and particle nature allows to define a vector $\vec{\chi}=\left(\chi^{a}_{0+},\chi^{a}_{0-},\chi^{a}_{1+},\chi^{a}_{1-},\chi^{p}_{0+},\chi^{p}_{0-},\chi^{p}_{1+},\chi^{p}_{1-} \right)$ where the indices are chosen as in Eq.~\eqref{eq:ansatz}. The space of functions is defined by
\begin{eqnarray}
e_i \in \left [  \frac{{\bf{k}}}{|{\bf{k}}|}  ,-\frac{{\bf{k}}}{|{\bf{k}}|} ,\frac{v_F{\bf{k}}}{T},\frac{v_F{\bf{k}}}{T}  ,\frac{{\bf{k}}}{|{\bf{k}}|},-\frac{{\bf{k}}}{|{\bf{k}}|},\frac{v_F{\bf{k}}}{T},\frac{v_F{\bf{k}}}{T}  \right]
\end{eqnarray}
 with $i=1,...,8$. One can expand the right hand side collision operator in Eq.~\eqref{eq:boltz} to linear order in the field $\bf{E}$, which leads to the following schematic expression
\begin{eqnarray}\label{eq:functional}
\frac{e {\bf{k}} \cdot{\bf{E}}}{T|{\bf{k}}|}f^{0,a}_+ \left( 1-f^{0,a}_+\right)&=& -\left( {\mathcal{C}}^{aa}+ {\mathcal{C}}^{ap}+{\mathcal{C}}_{\rm{dis}}^{aa}\right)\left [ \chi^{a/p}_{0/1,\pm} \right ]  \nonumber \\ -\frac{e {\bf{k}} \cdot{\bf{E}}}{T|{\bf{k}}|} f^{0,a}_- \left( 1-f^{0,a}_-\right)&=&-\left( {\mathcal{C}}^{aa} + {\mathcal{C}}^{ap}+{\mathcal{C}}_{\rm{dis}}^{aa} \right) \left [ \chi^{a/p}_{0/1,\pm} \right ]  \nonumber \\ 0&=& -\left( {\mathcal{C}}^{pp} + {\mathcal{C}}^{pa}+{\mathcal{C}}_{\rm{dis}}^{pp} \right) \left [ \chi^{a/p}_{0/1,\pm} \right ]  \nonumber \\ 0&=& -\left( {\mathcal{C}}^{pp} + {\mathcal{C}}^{pa}+{\mathcal{C}}_{\rm{dis}}^{pp} \right) \left [ \chi^{a/p}_{0/1,\pm} \right ]  \;. \nonumber \\
\end{eqnarray}

We define the scalar product between two objects in this space as 
\begin{eqnarray}
\langle a | b \rangle= \int \frac{d^2k}{(2 \pi)^2} a({\bf{k}}) b({\bf{k}})\;.
\end{eqnarray}
In the following we allow the more general case of applied fields in both layers. This is a generalization giving access to all quantities within the conductivity tensor including the passive layer conductivity, which in principle can be different from the active layer. 
We define a vector of the driving term as
\begin{eqnarray}\label{eq:drive}
\vec{D}&=&\left( \vec{D}_a,\vec{D}_p  \right) \; {\rm{with}}  \nonumber \\
\vec{D}_a&=&( \langle e_1 | D^a_+\rangle , \langle e_2 | D^a_-\rangle ,\langle e_3 | D^a_+\rangle ,\langle e_4 | D^a_-\rangle) \;,\nonumber \\
\vec{D}_p&=&( \langle e_5 | D^p_+\rangle , \langle e_6 | D^p_-\rangle ,\langle e_7 | D^p_+\rangle ,\langle e_8 | D^p_-\rangle) 
\end{eqnarray}
where 
\begin{eqnarray}
D^{a/p}_+&=&\frac{e v_F{\bf{k}} \cdot{\bf{E}}}{T|{\bf{k}}|}f^{0,a/p}_+ \left( 1-f^{0,a/p}_+\right) \; {\rm{ and}} \nonumber \\ D^{a/p}_-&=&-\frac{e v_F {\bf{k}} \cdot{\bf{E}}}{T|{\bf{k}}|} f^{0,a/p}_- \left( 1-f^{0,a/p}_-\right)\;.
\end{eqnarray}
Equivalently, the elements of the collision matrix are given by
\begin{eqnarray}\label{eq:collisionmatrix}
\hat{\mathcal{C}}_{ij}=\langle e_i | {\mathcal{C}}^{aa}+ {\mathcal{C}}^{ap}+{\mathcal{C}}_{\rm{dis}}^{aa}+ {\mathcal{C}}^{pp} + {\mathcal{C}}^{pa} +{\mathcal{C}}_{\rm{dis}}^{pp}  |e_j   \rangle 
\end{eqnarray}
where the superscripts on the right hand side indicate that the matrices act within layer space. 
Finally, the Boltzmann equation can be cast in the form
\begin{eqnarray}
\vec{D}=\hat{\mathcal{C}}\cdot {\vec{\chi}}\;.
\end{eqnarray}

A straightforward matrix inversion 
\begin{eqnarray}
\vec{\chi} =\hat{\mathcal{C}}^{-1} \cdot \vec{D}
\end{eqnarray}
allows to determine the expansion coefficients. The projection to obtain the respective single-layer conductivities and the transconductance is done via
\begin{eqnarray}
\vec{\chi}^a&=&\hat{\mathcal{C}}^{-1} \cdot (\vec{D}^a,0,0,0,0) \; {\rm and} \nonumber \\ \vec{\chi}^p&=&\hat{\mathcal{C}}^{-1} \cdot (0,0,0,0,\vec{D}^p) \;.
\end{eqnarray}

The conductivities in the individual layers now read
\begin{eqnarray}\label{eq:sa}
{\bf{\sigma}}^a=\frac{N\pi e^2}{hT}\sum_{i=1}^4 \vec{D}_i \vec{\chi}^a_i \big|_{|{\bf{E}}|=1}
\end{eqnarray}
and
\begin{eqnarray}
{\bf{\sigma}}^p=\frac{N\pi e^2}{hT}\sum_{i=1}^4 \vec{D}_{i+4} \vec{\chi}^p_{i+4} \big|_{|{\bf{E}}|=1}
\end{eqnarray}
while the off diagonal drag conductivity reads
\begin{eqnarray}\label{eq:sd}
{\bf{\sigma}}^d=\frac{N\pi e^2}{hT}\sum_{i=1}^4 \vec{D}_i \vec{\chi}^a_{i+4} \big|_{|{\bf{E}}|=1} \;,
\end{eqnarray}
where $\vec{D}$ was introduced in Eq.~\eqref{eq:drive}, and $N$ is the number of valley and spin degrees of freedom. We will give a concrete example of this formalism in a reduced setting in Sec.~\ref{sec:fcp} and Sec.~\ref{sec:fl}.

\section{Non-degenerate limit: $|\mu_{a/p}|/T \ll1$}\label{sec:nd}

In this limit the difference from the standard Fermi liquid behavior of Coulomb drag is expected to be largest: not only the transconductivity, but also the conductivity of the individual layers are expected to possibly be dominated by either inelastic or elastic scattering. This implies that we expect drastic changes as the ratio of disorder to interactions is changed. We will find that this ratio can alter the drag resistivity by orders of magnitude. In the discussion of this limit we neglect the effect of screening due to the lack of density of states, which is the rational given in Sec.~\ref{sec:sc}. Our analysis does not capture the regime of electron- and hole-puddles and we assume in the following that the temperature $T$ is high enough to be beyond the inhomogeneity scale. With increasing sample quality we expect that this regime can be pushed to rather low temperatures thereby increasing the domain of validity of our analysis. All calculations throughout this section have been performed explicitly for graphene, meaning $N=4$, meaning all numbers are calculated for this case. Since we are going behind large-N including crossed diagrams in our analysis we can not deduce the result for different values of $N$ by simply scaling.

\subsection{Both layers at the Dirac point}\label{sec:bd}
If both layers are at the Dirac point there is no drag due to particle-hole symmetry in both layers: Plugging the parametrization Eq.~\eqref{eq:lr} and Eq.~\eqref{eq:ansatz} into the Boltzmann equation of the passive layer it is straightforward but rather tedious to show that the equilibrium distribution function
\begin{eqnarray}
f^p_{\pm}(\mathbf{k},t) =f^{0,p}_{\pm}(v_{F}k)
\end{eqnarray}
solves the Boltzmann equation for the passive layer. This immediately implies that in the setup shown in Fig.~\ref{Fig:fig1} we have $V_2=R_2=0$, meaning the aforementioned absence of drag. 
However, in this limit we have a well-defined single-layer charge response even in the absence of disorder since the current carrying state in the active layer is of zero total momentum. This was found to be given by
\begin{eqnarray}\label{eq:sl}
\sigma_0=0.76 \frac{e^2}{h\alpha^2}
\end{eqnarray}
by a numerical solution of the Boltzmann equation in the leading logarithmic approximation~\cite{Fritz2008} with $\alpha \equiv \frac{e^{2}}{\varepsilon v_{F}}$ being the dimensionless fine structure constant. We note that a similar calculation taking into account only the large-N diagrams (discarding crossed diagrams) was carried out by Kashuba~\cite{Kashuba2008} and the result was $\sigma_0'=0.69 \frac{e^2}{h\alpha^2}$.
In the limit $Td/v_F \gg1$ we expect and find that the active layer conductivity extrapolates to this isolated single layer conductivity, $\sigma_0$. However, we find there is a substantial effect of the passive layer on the transport properties of the active layer in the limit $Td/v_F \leq \mathcal{O}(1)$. 

Overall, in the absence of screening effects the conductivity of the active layer (in the passive layer it is zero) is a one parameter function of the type
\begin{eqnarray}
\sigma_a(T,d)=\sigma_a \left(\frac{Td}{v_F}\right)\;.
\end{eqnarray}
We calculate this crossover function from a full numerical solution of the four coupled Boltzmann equations (at charge neutrality one could in principle reduce the number by a factor two since there is a particle-hole symmetry to exploit).
\begin{figure}[t]
\includegraphics[width=0.45\textwidth]{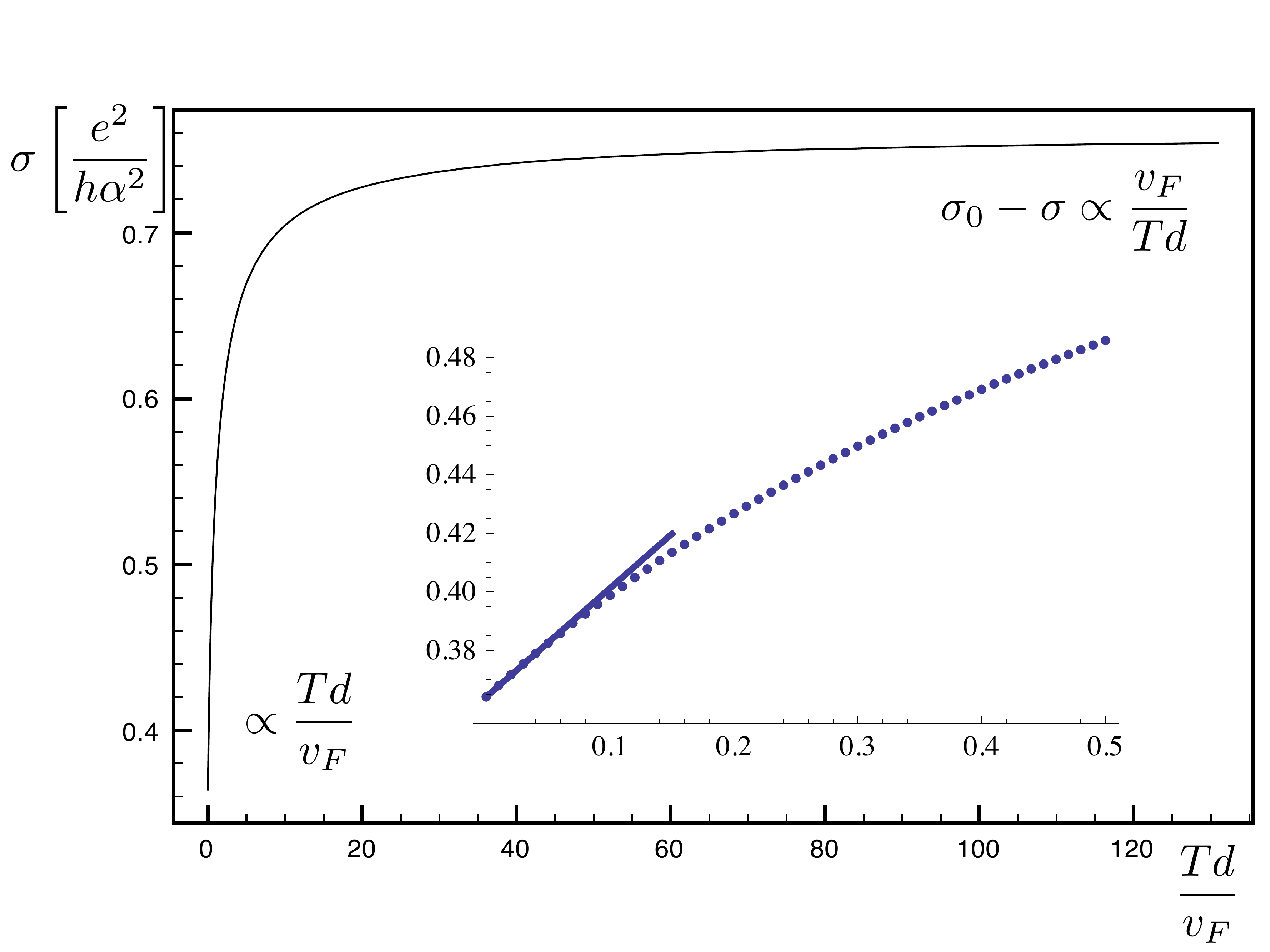}
\caption{Minimal d.c. conductivity $\sigma_a \left(\frac{Td}{v_F}\right)$ as a function of $T$ in units $\frac{v_F}{d}$. The conductivity interpolates between two limiting values at high and low temperatures and is described as a universal function which only depends upon $Td/v_F$. For low temperatures the behavior is given by $\lim_{Td/v_F \to 0}\left(\sigma \left ( Td/v_F\right)-0.36 \frac{e^2}{h\alpha^2}\right) \propto \frac{Td}{v_F} $ while for large temperatures it is given by $\lim_{Td/v_F \to \infty} \left(\sigma_0-\sigma \left( Td/v_F\right)\right)\propto \frac{v_F}{Td} $.}\label{Fig:fig2}
\end{figure}

It is shown in Fig.~\ref{Fig:fig2} as a function of the dimensionless parameter $Td/v_F$. It interpolates from $0.36\frac{e^2}{h\alpha^2}$ at low temperatures to $0.76\frac{e^2}{h\alpha^2}$ at high temperatures, which is the isolated single layer conductivity, see Eq.~\eqref{eq:sl}. This implies that for a fixed distance as a function of temperature the single layer conductivity can change by roughly a factor of two. 
We can easily rationalize the results found numerically in the limit $\frac{Td}{v_F} \to 0$: in the absence of charge currents between the layers the charge carriers in the active layer scatter from charge carriers in the active layer as well as from those in the passive layer. Both types of processes share the same Coulomb potential due to $\frac{Td}{v_F} \to 0$, which implies the exponential screening factor in Eq.~\eqref{eq:yukawa} is not active for typical momenta. The scattering times associated with inelastic scattering within and in-between layers will thus just add up. We can guess the result for the conductivity from the two single layer results mentioned above: For a clean sheet of graphene the conductivity~\cite{Fritz2008} assumes the value $\sigma_0=0.76 \frac{e^2}{h \alpha^2}$. This result takes into account all diagrams of the Born approximation, thus also crossed diagrams beyond large-N. In a calculation which discarded crossed diagrams and concentrated on large-N diagrams Kashuba~\cite{Kashuba2008} found the conductivity was given by $\sigma_0'=0.69 \frac{e^2}{h \alpha^2}$ for a single layer. Since scattering across layers only involves density-density type scattering (plasmons) it is faithfully accounted for by large-N type diagrams and we expect that in the limit $Td/v_F\ll1$ the conductivity is given by
\begin{eqnarray}
\sigma_a=\frac{1}{\frac{1}{\sigma_0}+\frac{1}{\sigma_0'}}\approx0.36 \frac{e^2}{h \alpha^2}\;,
\end{eqnarray}
which corresponds to adding the inverse scattering times.
The low temperature behavior can be rationalized from the exponential factor in the interlayer potential: the typical momentum of electrons and holes involved in the electronic transport is the thermal momentum $q_{\rm{typ}} = T/v_F$. Expanding the exponential of inter-layer Coulomb interaction for $\frac{Td}{v_F} \ll 1$ at this typical momentum thus yields a correction $U(q,T)\approx V(q,T)\left(1- q_{\rm{typ}} d \right)  = V(q,T)\left(1-  \frac{Td}{v_F}\right)$, which suggests a linear variation of the conductivity for very small temperatures. 

An interesting question is to speculate whether this effect should be visible in available geometries. Typical values of layer separations which are currently used in experiments are on the couple of nanometer range. For a distance of $d\approx 10$ nm this implies that the typical crossover temperature which we extract from demanding that $\frac{Td}{v_F} \approx 0.2-0.4$ is given by $T_{\rm{cross}}=\frac{v_F}{d} \approx 150K-300 K$. Consequently, in experiments one can easily reach the temperature range where interaction effects are enhanced. 

We note that we performed the same calculation by describing the passive layer in terms of plasmons. It was shown before that due to particle-hole symmetry this mode remains in equilibrium~\cite{Fritz2011} and thus one can calculate the scattering of electrons and holes from equilibrium bosons which simplifies the analysis. We have checked that the results are in perfect agreement.

\subsection{Increasing inelastic scattering close to the Dirac point}\label{sec:increased}

\begin{figure}
\begin{center}
\includegraphics[width=0.5\textwidth]{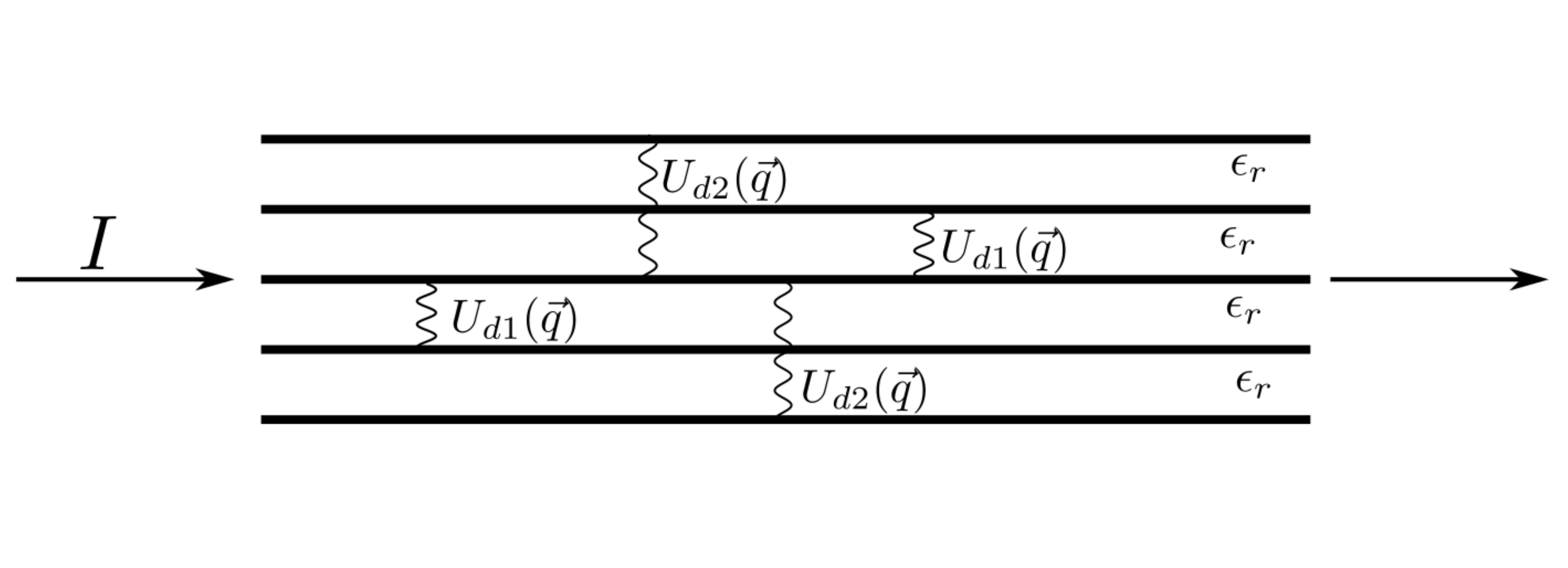}
\caption{Possible experimental setup which can effectively increase the role of inelastic scattering close to the Dirac point. The layers adjacent to the central active layer, in which a current $I$ is driven serve as reservoirs which provide a source of inelastic scattering. }\label{fig:layers}
\end{center}
\end{figure}

The above results immediately bring about a promising route towards increasing the effect of inelastic scattering in experiments carried out in the vicinity of the Dirac point in ultra clean samples: stacking a larger number of monolayers effectively increases the effect of inelastic scattering since the inverse scattering times due to interactions of the individual layers add up as long as $\frac{Td}{v_F} \ll 1$. Current experiments on bilayers have demonstrated that distances $d=1$ nm are conceivable without leakage currents~\cite{Ponomarenko} which implies that one could arrange a large number of layers in a sandwich structure and still be well below the crossover scale for temperatures up to the one hundred Kelvin range. A possible schematic setup is shown in Fig.~\ref{fig:layers} where the central layer is the active layer in which the current is driven while the surrounding passive layers solely increase the effect of interactions but remain in equilibrium themselves. We propose that such a sandwich structure can facilitate experiments in the limit of the sought after hydrodynamic interaction dominated regime~\cite{Mueller2008,Fritz2009}.

\subsection{Finite chemical potential}\label{sec:fcp}

In the following we concentrate on equal chemical potentials (our analysis trivially includes the situation of equal doping but with different types of charge carriers in the layers, meaning $\mu_a=-\mu_p$, which results in an overall minus sign) and comment on different chemical potentials in the individual layers in Sec.~\ref{sec:crossovers}.
Equal doping can be achieved in samples in which the charge carrier density can be controlled individually by separate gates. As discussed above, when both layers are at zero doping there is no drag. Even though for finite doping this is not true any more, the electron and hole density in this regime still is mainly thermal. Like in the case discussed in Sec.~\ref{sec:bd} we expect to find a strong interaction effect and enhanced inelastic scattering here as well, especially in the limit $\frac{T d}{v_F} \ll1$, where the interlayer interaction is essentially undamped.

The conclusions of the following discussions hold for arbitrary finite chemical potential but for small chemical potential we can carry out a simplified analysis. In this regime we can neglect the effect of screening and one can relate different parameters in terms of disorder and interaction strength easily by scaling the matrix elements. For simplicity, we also assume that in both layers there is the same amount of disorder. The collision matrix has the aforementioned block structure and assumes the form
\begin{widetext}
\begin{eqnarray}\label{eq:cm}
\hat{\mathcal{C}}=\alpha^2 \left(\begin{array}{cc}    \mathcal{C}^{aa}(\overline{\mu}) +  \mathcal{C}^{ap}(\overline{\mu},\overline{d})+\frac{1}{\overline{\alpha}^2}\mathcal{C}_{\rm{dis}} (\overline{\mu})&  -\mathcal{C}^{ap}(\overline{\mu},\overline{d}) \\  -\mathcal{C}^{ap}(\overline{\mu},\overline{d})  &   \mathcal{C}^{aa}(\overline{\mu}) +  \mathcal{C}^{ap}(\overline{\mu},\overline{d})+ \frac{1}{\overline{\alpha}^2}  \mathcal{C}_{\rm{dis}} (\overline{\mu}) \end{array} \right)=\alpha^2 \hat{\tilde{\mathcal{C}}}_{ij}
\end{eqnarray}
\end{widetext}
where $\overline{\alpha}^2=\frac{\alpha^2 T^2}{(Ze^2/\epsilon_r)^2\rho_{\rm{imp}}}$, $\overline{d}=\frac{Td}{v_F}$, and $\overline{\mu}=\frac{\mu}{T}$ are dimensionless parameters and $i,j$ denotes the layer indices $a$ and $p$. The dimensionless parameter $\overline{\alpha}^2$ corresponds to the ratio of elastic scattering to inelastic scattering, {\it i.e.}, $\tau_{\rm{imp}}/\tau_{ee}$. Furthermore, we have the driving terms
\begin{eqnarray}
\vec{D}_a (\overline{\mu})=\vec{D}_p (\overline{\mu}).
\end{eqnarray}
From this we find rather simple expression for the individual conductivities 
\begin{eqnarray}
\sigma_{a}(\alpha,\overline{\alpha},\overline{d},\overline{\mu}) &=&\sigma_{p}(\alpha,\overline{\alpha},\overline{d},\overline{\mu}) \nonumber \\ &=& \frac{N\pi e^2}{h\alpha^2}\vec{D}_a(\overline{\mu}) \cdot  \hat{\tilde{\mathcal{C}}}^{-1}_{aa} \cdot \vec{D}_a(\overline{\mu}) \; {\rm{and}}\nonumber \\ \sigma_{d}(\alpha,\overline{\alpha},\overline{d},\overline{\mu}) &=&  \frac{N \pi e^2}{h\alpha^2}\vec{D}_a(\overline{\mu}) \cdot   \hat{\tilde{\mathcal{C}}}^{-1}_{ap} \cdot \vec{D}_a(\overline{\mu}) 
\end{eqnarray}
leading to
\begin{eqnarray}\label{eq:dragen}
\rho_d&=& \frac{-\alpha^2\frac{h}{N\pi e^2}\vec{D}_a(\overline{\mu}) \cdot   \hat{\tilde{\mathcal{C}}}^{-1}_{ap} \cdot \vec{D}_a(\overline{\mu}) }{\left( \vec{D}_a(\overline{\mu}) \cdot  \hat{\tilde{\mathcal{C}}}^{-1}_{aa} \cdot \vec{D}_a(\overline{\mu})  \right)^2-\left(\vec{D}_a(\overline{\mu}) \cdot   \hat{\tilde{\mathcal{C}}}^{-1}_{ap} \cdot \vec{D}_a(\overline{\mu})  \right)^2}\nonumber \\ &=& \alpha^2 \frac{h}{N \pi e^2} g(\overline{\alpha},\overline{d},\overline{\mu})\;.
\end{eqnarray}
From the above collision matrix Eq.~\eqref{eq:cm} it is obvious that we have two limiting cases: $\overline{\alpha}=0$ corresponds to the disorder dominated limit, while $\overline{\alpha}\to \infty$ corresponds to the clean limit.

\noindent {\it The disorder dominated limit: $\overline{\alpha} \to 0$}

\noindent Using the collision matrix and its components introduced in Eq.~\eqref{eq:cm} it is straightforward to find that the leading order in $\overline{\alpha}$ expression of drag reads
\begin{eqnarray}
\rho_d \approx \alpha^2 \frac{h}{N \pi e^2} \frac{\vec{D}_a (\overline{\mu}) \cdot \mathcal{C}_{\rm{dis}}^{-1} (\overline{\mu}) \overline{\mathcal{C}}^{ap}(\overline{d},\overline{\mu})\mathcal{C}_{\rm{dis}}^{-1} (\overline{\mu})    \cdot \vec{D}_a (\overline{\mu})}{\left( \vec{D}_a (\overline{\mu}) \cdot \mathcal{C}_{\rm{dis}}^{-1} (\overline{\mu}) \cdot \vec{D}_a (\overline{\mu}) \right)^2}\;. \nonumber \\
\end{eqnarray}
This expression is well behaved and no singular matrix operations are involved. This is due to the fact that in the disordered limit the presence of impurities breaks translational invariance and the individual conductivities are always well defined. The above expression is consistent with the standard approximation 
\begin{eqnarray}
\rho_d=\frac{-\sigma_d}{\sigma_a \sigma_p-\sigma_d^2} \approx \frac{-\sigma_d}{\sigma_a \sigma_p} \;.
\end{eqnarray}
This approximation does not hold in the interaction dominated limit which in contrast to traditional two dimensional electronic systems might be attainable in graphene.
\noindent {\it The interaction dominated limit: $\overline{\alpha} \to \infty$}

\noindent For zero chemical potential the conductivity in all layers is well defined even in the absence of disorder due to particle-hole symmetry. However, drag also vanishes for the very same reason. This changes for finite chemical potential: in the limit of vanishing disorder the individual layer conductivities as well as the transconductivity diverge. However, it turns out that the drag resistance can still be finite. This is an effect of the boundary conditions which are such that no current is allowed to flow in the passive layer. The full collision matrix was introduced in Eq.~\eqref{eq:cm} and we see that in the limit of vanishing disorder ($\overline{\alpha}\to \infty$) it assumes a simplified form. At finite chemical potential it turns out that the matrix $\mathcal{C}^{aa}(\overline{\mu})$ is not invertible due to the existence of the momentum zero modes, which are excited. This is not true for $\mathcal{C}^{ap}(\overline{\mu},\overline{d})$ which is invertible. However, if $\mathcal{C}^{aa}(\overline{\mu})=0$ the full matrix is not invertible. This implies we have to take care performing the limit $\overline{\alpha}\to \infty$ and should not do so from the outset. Consequently, we need the effect of disorder acting on the zero modes of $\mathcal{C}^{aa}(\overline{\mu})$ in order to regularize the response within the individual layers. In order to isolate the space of zero modes we first transform the collision matrix by the matrix $U$ which diagonalizes the collision matrix $\mathcal{C}^{aa}(\overline{\mu})$, meaning we perform the operation
\begin{eqnarray}
\hat{\mathcal{C}}'=\left ( \begin{array} {cc}  U & 0 \\ 0  & U    \end{array}  \right)\hat{\mathcal{C}}\left ( \begin{array} {cc}  U^{-1} & 0 \\ 0  & U^{-1}    \end{array}  \right)
\end{eqnarray}
where $U$ is chosen such that
\begin{eqnarray}
U \mathcal{C}^{aa} (\overline{\mu}) U^{-1} = \mathcal{D}^{aa} (\overline{\mu})
\end{eqnarray}
where $\mathcal{D}^{aa}(\overline{\mu})$ is a diagonal matrix with zero eigenvalues corresponding to momentum conservation. In the following we will make the simplifying assumption (only for the purpose of concise presentation) that the sectors of the zero modes and the other modes do not mix. This implies we can carry out a simplified discussion of the individual conductivities. We introduce an index $0$ which refers to the the space of zero modes, and $1$ referring to the other components. The transformed collision matrix assumes the form (remember $\mathcal{D}^{aa}_{00}=0$)
\begin{widetext}
\begin{eqnarray}
\hat{\mathcal{C}}' = \alpha^2\left(\begin{array} {cccc}  \mathcal{C}_{00}^{ap} (\overline{\mu},\overline{d}) +  \frac{1}{\overline{\alpha}^2} \mathcal{C}_{00}^{\rm{dis}} (\overline{\mu}) &  0 &  -\mathcal{C}_{00}^{ap} (\overline{\mu},\overline{d}) & 0 \\   0 & \mathcal{D}_{11}^{aa}(\overline{\mu}) +\mathcal{C}_{11}^{ap} (\overline{\mu},\overline{d}) +  \frac{1}{\overline{\alpha}^2 }\mathcal{C}_{11}^{\rm{dis}} (\overline{\mu}) & 0 &  -\mathcal{C}_{11}^{ap} (\overline{\mu},\overline{d}) \\   -\mathcal{C}_{00}^{ap} (\overline{\mu},\overline{d})  & 0 &  \mathcal{C}_{00}^{ap} (\overline{\mu},\overline{d}) +  \frac{1}{\overline{\alpha}^2} \mathcal{C}_{00}^{\rm{dis}} (\overline{\mu}) & 0 \\ 0 &  -\mathcal{C}_{11}^{ap} (\overline{\mu},\overline{d})  & 0 & \mathcal{D}_{11}^{aa} (\overline{\mu})+\mathcal{C}_{11}^{ap} (\overline{\mu},\overline{d}) +  \frac{1}{\overline{\alpha}^2} \mathcal{C}_{11}^{\rm{dis}} (\overline{\mu})\end{array}   \right)\nonumber \\ 
\end{eqnarray} 
Using this expression one can first analyze the individual conductivities which read
\begin{eqnarray}
\sigma_a&=& \frac{N \pi e^2}{h  \alpha^2} \vec{D}_a (\overline{\mu}) U^{-1}\left( \begin{array}{cc} \frac{\overline{\alpha}^2}{2} \left( \mathcal{C}_{00}^{\rm{dis}}\right)^{-1}  & 0  \\   0  &  \left( \mathcal{D}_{11}^{aa}+\mathcal{C}_{11}^{ap}  -  \mathcal{C}_{11}^{ap} \left(\mathcal{D}_{11}^{aa}+\mathcal{C}_{11}^{ap}  \right)^{-1}  \mathcal{C}_{11}^{ap} \right)^{-1} \end{array} \right) U\vec{D}_a (\overline{\mu}) \nonumber \\ &=& \frac{N \pi e^2}{h  \alpha^2}\left(\vec{d}_a^0(\overline{\mu}),\vec{d}_a^1(\overline{\mu}) \right) \left( \begin{array}{cc} \frac{\overline{\alpha}^2}{2} \left( \mathcal{C}_{00}^{\rm{dis}}\right)^{-1}  & 0  \\   0  & \left( \mathcal{D}_{11}^{aa}+\mathcal{C}_{11}^{ap}  -  \mathcal{C}_{11}^{ap} \left(\mathcal{D}_{11}^{aa}+\mathcal{C}_{11}^{ap}  \right)^{-1}  \mathcal{C}_{11}^{ap}\right)^{-1}  \end{array} \right)\left( \begin{array}{c}  \vec{d}_a^0(\overline{\mu})  \\   \vec{d}_a^1(\overline{\mu}) \end{array}\right) \nonumber \\ 
\end{eqnarray}
and

\begin{eqnarray}
\sigma_d= -\frac{N \pi e^2}{h  \alpha^2}\left(\vec{d}_a^0(\overline{\mu}),\vec{d}_a^1(\overline{\mu}) \right)\left( \begin{array}{cc} \frac{\overline{\alpha}^2}{2} \left( \mathcal{C}_{00}^{\rm{dis}} \right)^{-1}  & 0  \\   0  &  \left(\mathcal{D}_{11}^{aa}+\mathcal{C}_{11}^{ap}  -  \mathcal{C}_{11}^{ap} \left(\mathcal{D}_{11}^{aa}+\mathcal{C}_{11}^{ap}  \right)^{-1}  \mathcal{C}_{11}^{ap} \right)^{-1} \mathcal{C}^{ap}_{11} \left(\mathcal{D}^{aa}_{11} +\mathcal{C}^{ap}_{11}\right)^{-1}  \end{array} \right) \left( \begin{array}{c}  \vec{d}_a^0(\overline{\mu} ) \\   \vec{d}_a^1(\overline{\mu}) \end{array}\right) \;. \nonumber \\ \end{eqnarray}
We can now understand why the conductivities diverge. In the case of interlayer conductivity $\sigma_a$ there now is one contribution $\propto \vec{d}_a^0(\overline{\mu}) \frac{\overline{\alpha}^2}{2} \left( \mathcal{C}_{00}^{\rm{dis}}\right)^{-1}   \vec{d}_a^0(\overline{\mu})$ which diverges in the limit $\overline{\alpha} \to \infty$, as it should. The very same term is responsible for the divergence of $\sigma_d$. Despite these divergencies the transresistivity remains finite, since the most severe divergences cancel. We have done this explicitly and found that
\begin{eqnarray}
\lim_{\overline{\alpha}\to \infty}\rho_d=-\alpha^2 \frac{2h}{N \pi e^2} \frac{1}{\vec{d}_a^1(\overline{\mu})   \left(\mathcal{D}_{11}^{aa}+\mathcal{C}_{11}^{ap}  -  \mathcal{C}_{11}^{ap} \left(\mathcal{D}_{11}^{aa}+\mathcal{C}_{11}^{ap}  \right)^{-1}  \mathcal{C}_{11}^{ap} \right)^{-1}\left (\mathbb{I} -\mathcal{C}^{ap}_{11} \left(\mathcal{D}^{aa}_{11} +\mathcal{C}^{ap}_{11}\right)^{-1}\right) \vec{d}_a^1(\overline{\mu})} \;,
\end{eqnarray}
\end{widetext}

which indeed is finite in the limit $\overline{\alpha}\to\infty$ since no singular matrices are involved and all information about disorder is gone. However, the validity of the above expression is restricted to finite values of $\overline{d}$, which becomes apparent from the fact that if $\mathcal{C}^{ap} \to 0$ in the limit $\overline{d}\to \infty$ it remains finite. This is rooted in the implicit assumption of the above analysis that $\mathcal{C}_{00}^{ap} \gg \frac{1}{\overline{\alpha}^2}\mathcal{C}^{\rm{dis}}$, which does not hold in the limit $d\to \infty$. The above finiteness of the response is similar to the so-called universal conductivity, where the Kubo expression is regularized by finite disorder, which drops out in final expression, allowing to extrapolate to the clean limit.

\begin{figure}[h!]
\includegraphics[width=0.45\textwidth]{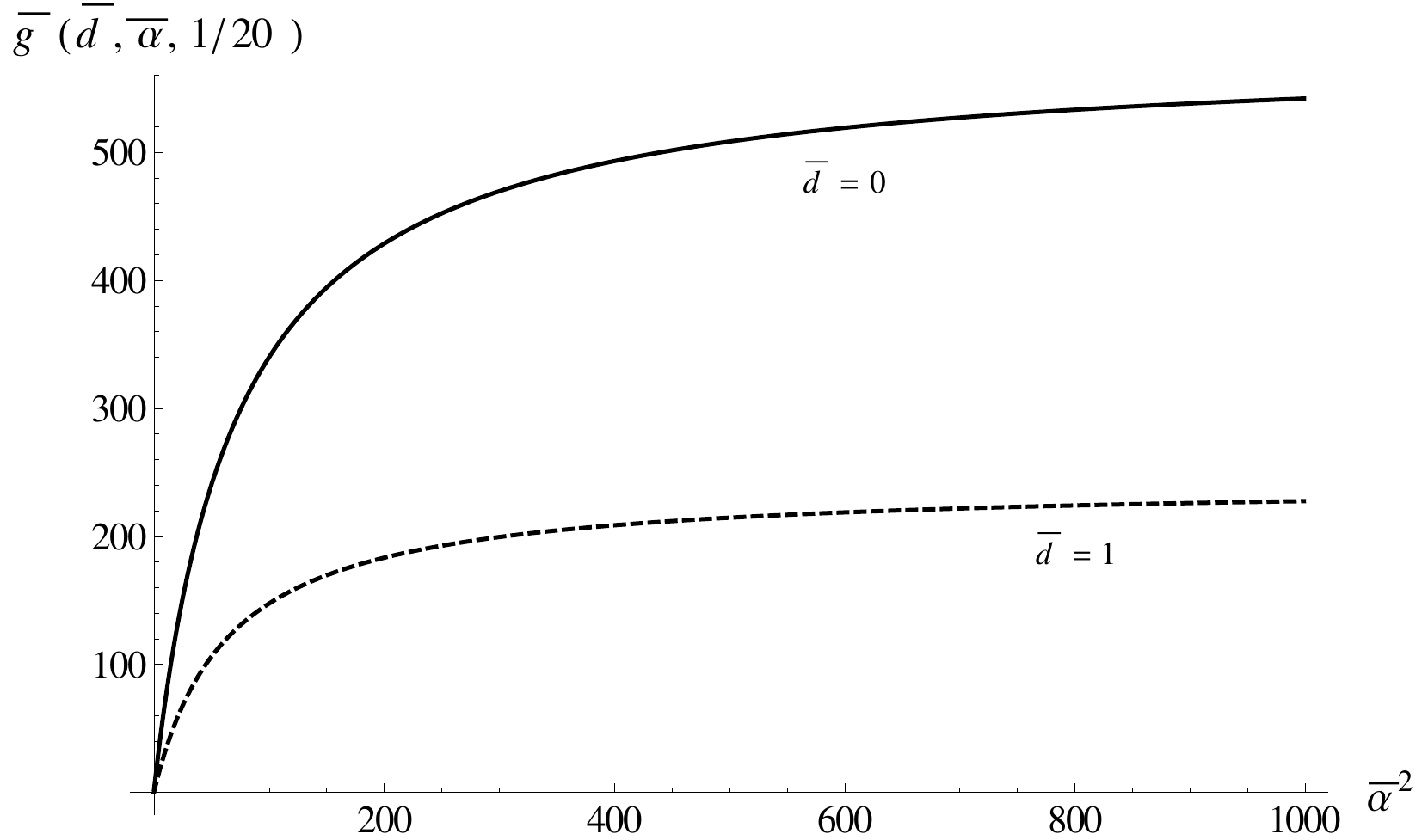}
\caption{Crossover function describing drag all the way from the disorder limited ($\overline{\alpha}\to 0$) to the clean system $\overline{\alpha} \to \infty$ for different dimensionless distances $\overline{d}=0,1$ and $\overline{\mu}=1/20$. Importantly this quantity saturates which we have checked explicitly for different realizations of $\overline{\mu}$.}\label{fig:crossovera}
\end{figure}

Overall, we have shown that drag resistivity is indeed described by a function of the type Eq.~\eqref{eq:dragen} which is always finite
\begin{eqnarray}
0<| g(\overline{\alpha}=\infty,\overline{d}<\infty,\overline{\mu})|  <\infty\;,
\end{eqnarray}
meaning the drag resistivity remains finite even in a clean system (strictly speaking we used $\overline{\mu}\ll1$ to discard screening without loss of generality). 
For very small values of $\mu/T$ we expect the drag to be $\propto (\mu/T)^2$ for symmetry reason which is also backed up by our numerical analysis. We studied drag in the vicinity of the Dirac point as a function of the dimensionless parameter $\overline{\alpha}$ in great detail. The clean system is found in the limit $\overline{\alpha}\to \infty$. A first observation is that for finite chemical potential we can make the extrapolation to the clean system and find that the drag resistivity remains finite. This can be seen in Fig.~\ref{fig:crossovera} where for distances $\overline{d}=0,1$ and for $\mu/T=1/20$ we have plotted the function
\begin{eqnarray}
\overline{g}(\overline{\alpha},\overline{d},\overline{\mu})=1/\overline{\mu}^2 g(\overline{\alpha},\overline{d},\overline{\mu})
\end{eqnarray}
 as a function of $\overline{\alpha}$. We have checked that this extrapolation can be performed for any finite chemical potential and the limiting value in the clean system, $\overline{\alpha}\to \infty$, increases upon decreasing the chemical potential.

\begin{figure}[h!]
\includegraphics[width=0.45\textwidth]{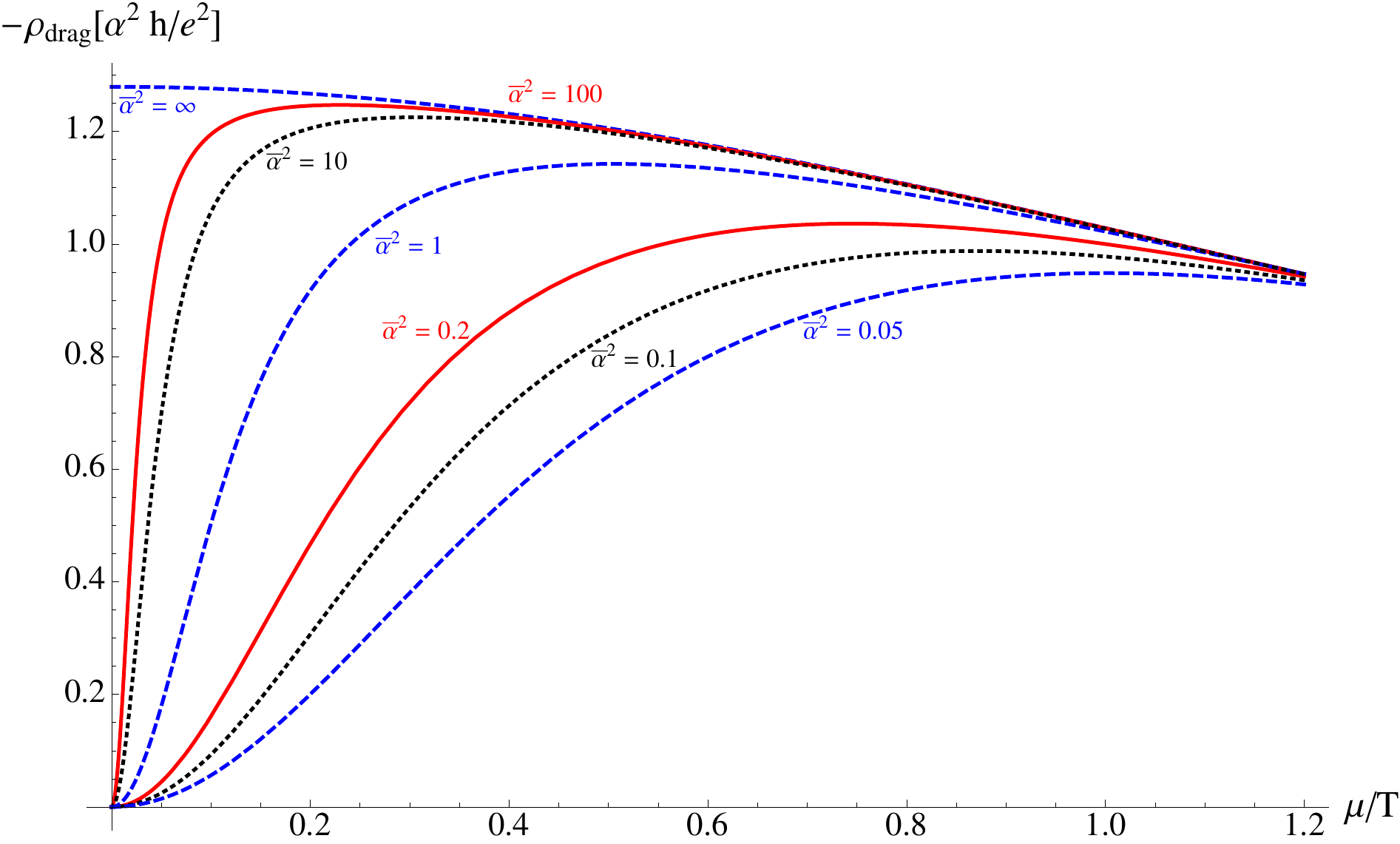}
\caption{Drag as a function of $\mu/T$ for different disorder strengths ranging from the disorder limited ($\overline{\alpha}\to 0$) all the way to the clean system $\overline{\alpha} \to \infty$. We find that in the limit $\mu\to 0$ we end a with a finite drag upon extrapolation. The value to which $\rho_d$ extrapolates is given by $\rho_d=-\frac{1}{\sigma_0}$ where $\sigma_0$ is the single layer conductivity and was defined in Eq.~\eqref{eq:sl}.}\label{fig:crossovers}
\end{figure}

This is in contrast to the statement based on particle-hole symmetry that drag at the Dirac point is zero. This signals that in the limit $\overline{\alpha} \to \infty$, which corresponds to the ballistic system, extrapolating to zero density must become singular which we show explicitly. We have furthermore studied the drag resistivity as a function of $\mu/T$ for a set of different disorder realizations, meaning different values of the parameter $\overline{\alpha}$. For finite disorder the overall shape is such that there is a maximum of drag resistivity.  In the limit of very low chemical potentials compared to the maximum position we indeed find behavior of the $(\mu/T)^2$ type which is consistent with expectations based on symmetry considerations. We find that upon decreasing disorder the maximum of drag shifts towards lower chemical potentials and pushes to zero in the clean limit. Consequently, with decreasing disorder the quadratic regime becomes increasingly small and it vanishes in the limit of zero disorder. These behaviors are extracted from our numerical results which are summarized in Fig.~\ref{fig:crossovers} where again we have discarded the role of screening and all the curves are plotted for $d=0$. 

\begin{figure}[h!]
\includegraphics[width=0.45\textwidth]{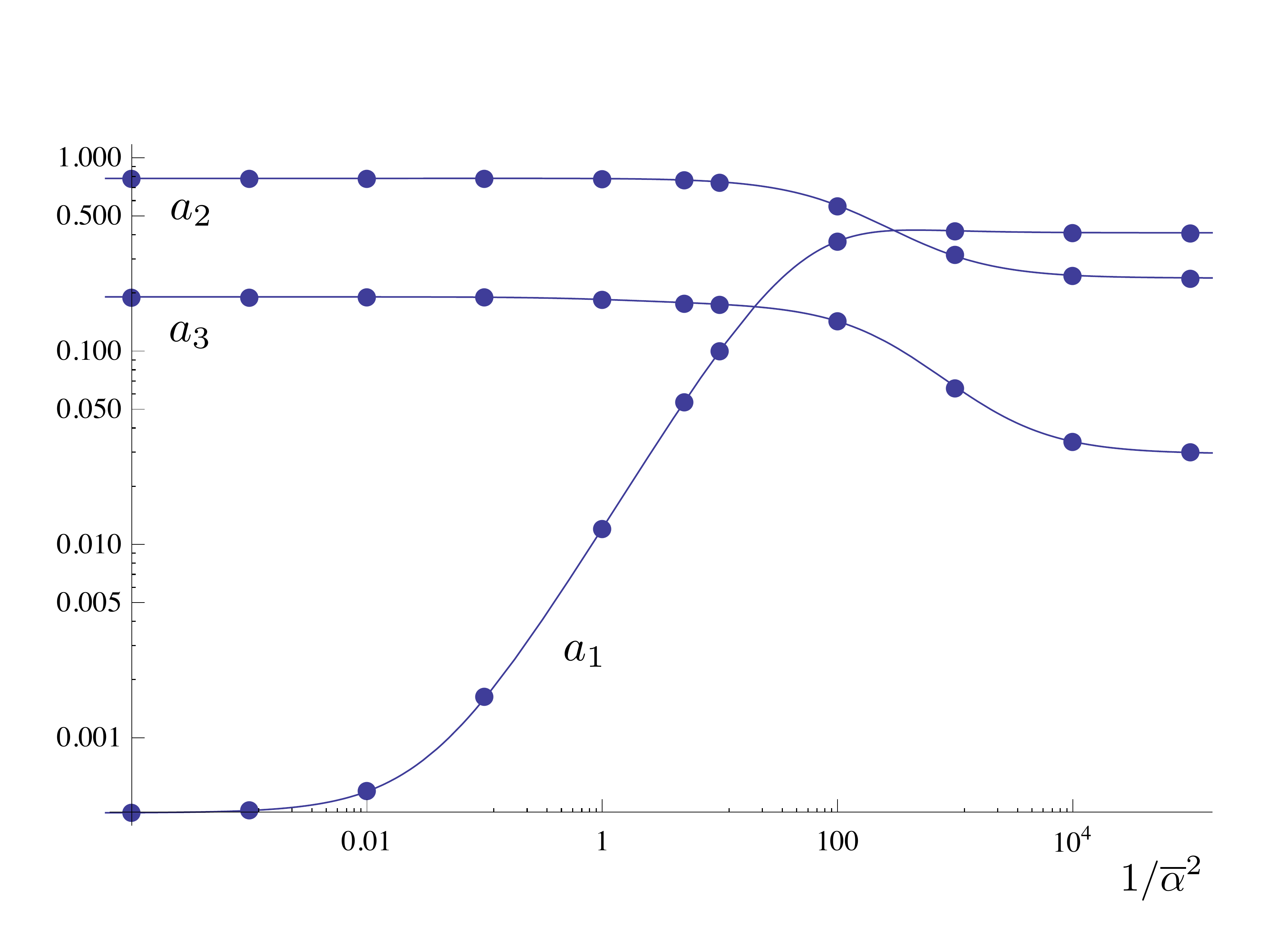}
\caption{Fitting parameters $a_1-a_3$ as a function of the parameter $1/\overline{\alpha}^2$. Zero corresponds to the clean limit while large values correspond to the dirty limit. Most importantly, to within numerical accuracy we find that $a_1\to 0$ as $\overline{\alpha}\to \infty$. This implies that in principle in the clean system at charge neutrality there can be finite drag.}\label{fig:fit}
\end{figure}

The extrapolation to $\overline{\alpha}=\infty$, {\it i.e.}, the clean case requires some care since it is very sensitive to small numerical errors. We show below how we obtained the limiting curve in Fig.~\ref{fig:crossovers}. We extract this behavior from fitting the numerical curves for different disorder realizations and extrapolating to the clean limit. In order to do so we have fitted the ascent of the curves with the following fitting curve
\begin{eqnarray}
-\rho_d/(\alpha^2 h/e^2) = \frac{\overline{\mu}^2}{a_1 (\overline{\alpha}^2)+a_2 (\overline{\alpha}^2) \overline{\mu}^2+a_3 (\overline{\alpha}^2) \overline{\mu}^4}\;.
\end{eqnarray}
The results of this fitting procedure for $a_1 (\overline{\alpha}^2)$, $a_2 (\overline{\alpha}^2)$,  $a_3 (\overline{\alpha}^2)$ and  are shown in Fig.~\ref{fig:fit}. 

Most importantly, we find that in the clean limit, $a_1 \to 0$ within our numerical accuracy, while $a_2$ and $a_3$ remain finite. In the limits of $\mu/T \ll1$ and $1/\overline{\alpha}^2 \ll 1$ we find
\begin{eqnarray}
\rho_d\left(\overline{\mu} \ll1,\frac{1}{{\overline{\alpha}^2}}\ll1 \right) = - \frac{\alpha^2 h}{e^2}\frac{\overline{\mu}^2}{0.76 \overline{\mu}^2 + 0.012 \frac{1}{
\overline{\a}^{2}}}\;.
\end{eqnarray}

This implies that upon performing the limit $\overline{\mu}\to 0$ after performing the extrapolation to the clean limit, $\overline{\alpha}\to \infty$, we end up with finite drag. Consequently, the drag resistivity depends upon the order of limits according to
\begin{eqnarray}
\lim_{\mu_a=\mu_p \to 0} \lim_{\overline{\alpha}\to \infty} \rho_d&=&-\frac{1}{\sigma_0} \nonumber \\
 \lim_{\overline{\alpha}\to \infty} \lim_{\mu_a=\mu_p \to 0} \rho_d&=&0 \;,
\end{eqnarray}
where $\sigma_0$ was introduced in Eq.~\eqref{eq:sl} and denotes the single layer conductivity. A finite limiting value has already been observed in the recent  work by Sch\"utt {\it et al.}~\cite{schuett2012}. Interestingly, studying their numbers shows that their result is consistent with the statement $\lim_{\mu_a=\mu_p \to 0} \lim_{\overline{\alpha}\to \infty}\rho_d=-\frac{1}{\sigma_0}$ even though there seem to be numerical discrepancies. Sch\"utt {\it et al.} worked in the framework of the large-$N$ approximation, in which crossed diagrams are neglected as opposed to our analysis.  Within that approximation the single layer conductivity has to be replaced by $\sigma_0'=0.69\frac{e^2}{h\alpha^2}$~\cite{Kashuba2008} which is consistent with the numerical value found. We thus conclude by saying that our results are not only qualitatively but also quantitatively compatible taking into account the slightly different approximation schemes used.
 
It is important to point out that in the physical system where $\overline{\alpha}$ is finite the drag at the Dirac point is always zero for symmetry reasons. We note that there still is a way towards finite drag at the Dirac point which is rooted in including $\alpha^3$ processes which however is beyond our scope~\cite{levchenko2008,schuett2012}. In that case the symmetry arguments ensuring zero drag are invalid and finding finite drag at the Dirac point is possible. To summarize, our result suggests that for extremely clean samples in the ballistic limit it the regime of doping in which the drag resistivity goes to zero can in principle becoming very narrow. However, we stress that we do not think that this effect is at the heart of the experimentally observed zero-bias drag in graphene~\cite{geim2012}. For any finite disorder level the fact that the drag resistivity drops to zero at the Dirac point seems inevitable to the order we consider here. More likely it is rooted in the $\alpha^3$-contribution~\cite{schuett2012} or related to a mechanism which relies on energy transfer between the two inhomogeneous layers~\cite{levitov2012}.

\section{Fermi liquid regime: $\mu/T \gg1$}\label{sec:fl}

In the Fermi liquid regime screening effects become crucial, see Sec.~\ref{sec:sc}. In the limit of strong doping it is reasonable to consider in both layers only one species of charge carrier. We assume that the chemical potential in both layers is large and positive implying that only electrons are involved in the processes. Since we are in the disorder dominated regime we restrict our analysis to the momentum mode implying that instead of working with matrices of dimension eight we can work with matrices of dimension two. We have checked that this reduction in the Fermi liquid regime leads to numerically identical results with the calculation involving all $64$ matrix elements. Again, both layers are characterized by identical charge carrier concentration as well as disorder level. The reduced Boltzmann equation reads
\begin{widetext}
\begin{eqnarray}
\left( \begin{array} {c} \langle e_3| D^a_+ \rangle \\ 0   \end{array}	  \right)= \left (\begin{array} {cc} \langle e_3| \mathcal{C}^{aa}+\mathcal{C}^{ap}+\mathcal{C}^{aa}_{\rm{dis}} |e_3 \rangle  &  \langle e_3| \mathcal{C}^{ap} |e_7 \rangle  \\  \langle e_7| \mathcal{C}^{pa}|e_3 \rangle  &  \langle e_7|\mathcal{C}^{pp}+\mathcal{C}^{pa}+\mathcal{C}^{pp}_{\rm{dis}} |e_7 \rangle \end{array} \right)\cdot \left(\begin{array}{c} \chi^a_{1,+}\\ \chi^p_{1,+} \end{array} \right)\;.  \nonumber \\ 
\end{eqnarray}
\end{widetext}
For reasons of momentum conservation there is no relaxation of the current due to intralayer interactions, which implies that $\langle e_3|\mathcal{C}^{aa}|e_3 \rangle=\langle e_7|\mathcal{C}^{pp}|e_7 \rangle=0$. Defining $\langle e_3| \mathcal{C}^{ap} |e_7 \rangle= -\mathcal{C}_{\rm{Coul}}$ this implies that $\langle e_7| \mathcal{C}^{pa} |e_3 \rangle= -\mathcal{C}_{\rm{Coul}}$, while $\langle e_3| \mathcal{C}^{ap} |e_3 \rangle =\langle e_7| \mathcal{C}^{pa} |e_7 \rangle= \mathcal{C}_{\rm{Coul}}$ for symmetry reasons, consistent with an infinite response in absence of impurities. Furthermore, we choose $\langle e_3| \mathcal{C}^{aa}_{\rm{dis}} |e_3 \rangle =\langle e_7|\mathcal{C}^{pp}_{\rm{dis}} |e_7 \rangle = \mathcal{C}_{\rm{dis}}$ and introduce the shorthand $\mathcal{D}=\langle e_3| D^a_+ \rangle$.
The individual conductivities read
\begin{eqnarray}
\sigma_a&=& \frac{N\pi e^2}{hT}  \frac{\mathcal{D}^2 (\mathcal{C}_{\rm{dis}}+\mathcal{C}_{\rm{Coul}})}{\mathcal{C}_{\rm{dis}}^2+2\;\mathcal{C}_{\rm{Coul}}\mathcal{C}_{\rm{dis}}} \;,  \nonumber \\ \sigma_p &=&  \frac{N\pi e^2}{hT}  \frac{\mathcal{D}^2 (\mathcal{C}_{\rm{dis}}+\mathcal{C}_{\rm{Coul}})}{\mathcal{C}_{\rm{dis}}^2+2\;\mathcal{C}_{\rm{Coul}}\mathcal{C}_{\rm{dis}}} \;,  \; {\rm{and}} \nonumber \\ \sigma_d&=& \frac{N\pi e^2}{hT}  \frac{\mathcal{D}^2 \mathcal{C}_{\rm{Coul}}}{\mathcal{C}_{\rm{dis}}^2+2\;\mathcal{C}_{\rm{Coul}}\mathcal{C}_{\rm{dis}}}  \ \;. 
\end{eqnarray}

From this the drag resistivity obtains as
\begin{eqnarray}
\rho_d =- \frac{hT}{N\pi e^2}\frac{\mathcal{C}_{\rm{Coul}}}{\mathcal{D}^2} \;.
\end{eqnarray}
We stress that no further approximation has been used to arrive at this final result.
The driving term is given by
\begin{eqnarray}
  \mathcal{D} \left( \frac{\mu}{T} \gg 1  \right) = \frac{v_F^2}{2\pi T^2} \left( \frac{\mu}{T} \right)^2 \left( \frac{T}{v_F}\right)^3
\end{eqnarray}
and $\mathcal{C}_{\rm{Coul}}$ assumes the relatively simple form 
\begin{eqnarray}\label{eq:coulombfermi}
\mathcal{C}_{\rm{Coul}}&=&\frac{4 N v_F^2}{T^4}\int \frac{d^2k}{(2\pi)^2}\frac{d^2k_1}{(2\pi)^2}\frac{d^2q}{(2\pi)^2}\times \nonumber \\ &\times& 2\pi\delta \left(v_Fk+v_Fk_1-v_F|{\bf{k}}+{\bf{q}}|-v_F|{\bf{k}}_1-{\bf{q}}| \right) \times \nonumber \\  &\times&{\bf{q}}\cdot {\bf{q}} |\tilde{T}_{++++}(\mathbf{k},\mathbf{k_{1}},\mathbf{q})|^{2} \times\nonumber \\ &\times& f^0_+(k)f^0_+(k_1)(1-f^0_+(|{\bf{k}}+{\bf{q}}|))(1-f^0_+(|{\bf{k}}_1-{\bf{q}}|))\;,\nonumber \\
\end{eqnarray}
where $ \tilde{T}_{++++} $ was defined in Eq.~\eqref{eq:tmatrix}. This leads to the following expression for the drag resistivity
\begin{widetext}
\begin{eqnarray}\label{eq:draganalytical}
\rho_d(\mu/T \gg1 )&=& -\frac{h}{  e^2} \left(\frac{T}{\mu} \right)^4 \frac{2 \pi^2 v_F^4}{T^5}   \int \frac{d^2k}{(2\pi)^2}\frac{d^2k_1}{(2\pi)^2}\frac{d^2q}{(2\pi)^2}\delta \left(v_Fk+v_Fk_1-v_F|{\bf{k}}+{\bf{q}}|-v_F|{\bf{k}}_1-{\bf{q}}| \right){\bf{q}}\cdot {\bf{q}}  \nonumber \\  &\times& |U_{ap}({\bf{q}},v_F(k-|{\bf{k}}+{\bf{q}}|))|^2  \Bigg| \left(1+\frac{(K+Q)^*K}{k|{\bf{k}}+{\bf{q}}|}   \right)  \left(1+\frac{(K_1-Q)^*K_1}{k_1|{\bf{k}}_1-{\bf{q}}|}   \right) \Bigg|^2 \nonumber \\ &\times& f^0_+(k)f^0_+(k_1)(1-f^0_+(|{\bf{k}}+{\bf{q}}|))(1-f^0_+(|{\bf{k}}_1-{\bf{q}}|))\;.
\end{eqnarray}
\end{widetext}

We continue to show how our approach recovers the standard formulae of drag~\cite{kamenev1995,flensberg1995} in a Fermi liquid in a straightforward manner, which is usually derived in the framework of a Kubo formula calculation using Ward identities. There are two key rewritings that we use in the following. We use the identity

\begin{eqnarray}\label{eq:delta}
&&\delta(v_Fk+v_Fk_1-v_F|{\bf{k}}+{\bf{q}}|-v_F |{\bf{k}}_1-{\bf{q}}|)=\nonumber \\
&& \int d \omega \delta (\omega-v_F k+v_F |{\bf{k}}+{\bf{q}}|)\delta (\omega+v_F k_1-v_F |{\bf{k}}_1-{\bf{q}}|) \nonumber \\
\end{eqnarray}
as well the identity between Bose and Fermi functions 
\begin{eqnarray}
&&f^0_+(v_Fk)(1-f^0_+(v_Fk-\omega)) =\nonumber \\
&&n_B(\omega)\left(f^0_+(v_Fk)-f^0_+(v_F k-\omega) \right)\;.
\end{eqnarray}
Both expressions have to be used twice for $k$ and $k_1$ separately. In this expression we use
\begin{eqnarray}
n_B(\omega)=\frac{1}{e^{\frac{\omega}{T}} -1}
\end{eqnarray}
which is the standard Bose function. The  imaginary part of the retarded polarization function can be written as
\begin{widetext}
\begin{eqnarray}
\operatorname{Im} \Pi^{++}_{a,p} ({\bf{q}},\omega)= N \pi  \int \frac{d^2k}{(2\pi)^2}  \delta (\omega-v_F k+v_F |{\bf{k}}+{\bf{q}}|)\frac{1}{4}\Bigg| \left(1+\frac{(K+Q)^*K}{k|{\bf{k}}+{\bf{q}}|}   \right)  \Bigg|^2( f^0_+(v_Fk)-f^0_+(v_F |{\bf{k}}+{\bf{q}}|)) )
\end{eqnarray}
\end{widetext}
where the superscript $++$ signals we only consider the electronic part (or $--$ for the hole part). This becomes asymptotically exact in the limit $\mu/T \to \infty$ which is what we concentrate on. Plugging these expressions into Eq.~\eqref{eq:draganalytical} and using 
\begin{eqnarray}
n_B(\omega) n_B(- \omega) = - \frac{1}{4 \sinh^2 \frac{\omega}{2T}}
\end{eqnarray}
we obtain the well-known formula~\cite{kamenev1995,flensberg1995,narozhny2011,polini2012}
\begin{widetext}
\begin{eqnarray}\label{eq:rhodragfl}
\rho_d(\mu/T \gg1 )&=&  \frac{h}{e^2} \frac{2}{\pi^2} \frac{1}{\mu_a \mu_p \nu_a \nu_p} \frac{1}{T}  \int d\omega \int \frac{d^2q}{(2\pi)^2} \frac{q^2}{\sinh^2 \frac{\omega}{2T}} |U_{ap}({\bf{q}},\omega))|^2 \operatorname{Im} \Pi^{++}_{p} ({\bf{q}},\omega) \operatorname{Im} \Pi^{++}_{a} (-{\bf{q}},-\omega)   \;,
\end{eqnarray}
\end{widetext}
where $ \nu_{a,p} $ is the density of states at the Fermi level, which in graphene is given by $ \nu_{a,p} = \frac{N \mu_{a,p}}{2 \pi v_F^2}$. The limiting values of the transresistivity can now be estimated upon knowledge of the screened interaction potential which according to Eq.~\eqref{eq:RPA} is given by
\begin{eqnarray}
U_{ap}({\bf{q}},\omega))=\frac{U}{(1+V \Pi_a)(1+V\Pi_p)-U^2 \Pi_a \Pi_p}\;.
\end{eqnarray}

In the ballistic limit the polarization function reads
\begin{eqnarray}\label{eq:polarizationfl}
 \operatorname{Im} \Pi^{++}_{a,p} (q,\omega,\mu/T \gg1 ) =  \nu_{a,p} \frac{\omega}{v_F q} \Theta (v_F q - |\omega|)\;.
\end{eqnarray}
Furthermore, we use Eq.~\eqref{eq:screen} to describe the effect of screening.

Rescaling variables to $ q \to \frac{q T}{v_F} $ and  $ \omega \to \omega T $ yields (we use $\overline{d}=\frac{Td}{v_F}$) and performing a number of manipulations we can rewrite the drag resistivity as
\begin{eqnarray}\label{eq:drag}
\rho_d  = \frac{h}{e^2} \left( \frac{T}{\alpha \mu} \right)^2 \frac{16 \pi}{N^4} \frac{1}{\left(k_F d \right)^4} g(\overline{d},\kappa)\;,
\end{eqnarray}
where we used $k_F=\frac{\mu}{v_F}$ and introduced the shorthand $\kappa=\frac{2\pi}{\alpha N k_F d}$. Furthermore, we introduced the function
\begin{eqnarray}
g(\overline{d},\kappa)=\int_0^\infty  \frac{dq q^3 e^{-2q}}{\left( \left(q \kappa+1 \right)^2-e^{-2q}\right)^2} \int_{-q/\overline{d}}^{q/\overline{d}}  \frac{d \omega\omega^2}{\sinh \frac{\omega}{2}}\;. 
\end{eqnarray}
As mentioned before, we are considering the limit $\mu/T \gg1$, but instead we leave $\overline{d}$ as well as $k_F d$ arbitrary. This implies there are different regimes, which one can access in the above formula. We have checked that all results from Eq.~\eqref{eq:rhodragfl} are numerically identical to the ones obtained from directly integrating Eq.~\eqref{eq:draganalytical}.

\subsection{Drag in the limit $\overline{d} \ll 1$}

This is the limit in which we expect to recover the standard results of Fermi liquid theory. We can approximate the function
\begin{eqnarray}
g(\overline{d}\ll 1,\kappa) &\approx& \int_0^\infty  \frac{dq q^3 e^{-2q}}{\left( \left(q \kappa+1 \right)^2-e^{-2q}\right)^2}   \int_{-\infty}^{\infty}  \frac{d \omega\omega^2}{\sinh \frac{\omega}{2}} \nonumber \\ &=& \frac{8 \pi^2}{3}  \int_0^\infty  \frac{dq q^3 e^{-2q}}{\left( \left(q \kappa+1 \right)^2-e^{-2q}\right)^2}  
\end{eqnarray}
implying it is independent of $T$ and thus drag has the standard $T^2$-behavior. This expression has two limiting behaviors as a function $k_F d$.

\noindent
{\it $k_F d \ll 1$:}

\noindent
In this limit, we find that the role of $\kappa$ cannot be neglected and consequently we have
\begin{eqnarray}
g(\overline{d} \ll 1,\kappa \gg 1) &\approx& \frac{8 \pi^2}{3} \frac{1}{\kappa^4} \int_{1/\kappa}^\infty  \frac{dq  e^{-2q}}{q}  \nonumber \\ &\approx& -\frac{8\pi^2}{3} \frac{\ln \kappa}{\kappa^4}
\end{eqnarray}
This implies the drag resistivity reads
\begin{eqnarray}
\rho_d \approx  \frac{h}{e^2} \left ( \frac{T}{\mu}\right)^2 \frac{8 \alpha^2}{3 \pi} \ln \frac{\alpha N k_F d}{2\pi} \;.
\end{eqnarray}

\noindent
{\it $k_F d \gg 1$:}

\noindent
In this limit, we find that the role of $\kappa$ can be neglected and consequently we have
\begin{eqnarray}
g(\overline{d} \ll 1,\kappa \gg 1) &\approx&  \frac{8 \pi^2}{3}  \int_0^\infty  \frac{dq q^3 e^{-2q}}{\left(1-e^{-2q}\right)^2}  \nonumber \\ &=& \pi^2 \zeta (3)\;,
\end{eqnarray}
where $\zeta(x)$ is the Riemann function and $\zeta(3) \approx 1.202$. 
This implies the drag resistivity reads
\begin{eqnarray}
\rho_d \approx \frac{h}{e^2} \left ( \frac{T}{\mu}\right)^2 \frac{16 \pi^3 \zeta(3)}{N^4} \frac{1}{\left( k_F d\right)^4 \alpha^2}\;.
\end{eqnarray}

\begin{figure}[h]
\includegraphics[width=0.49\textwidth]{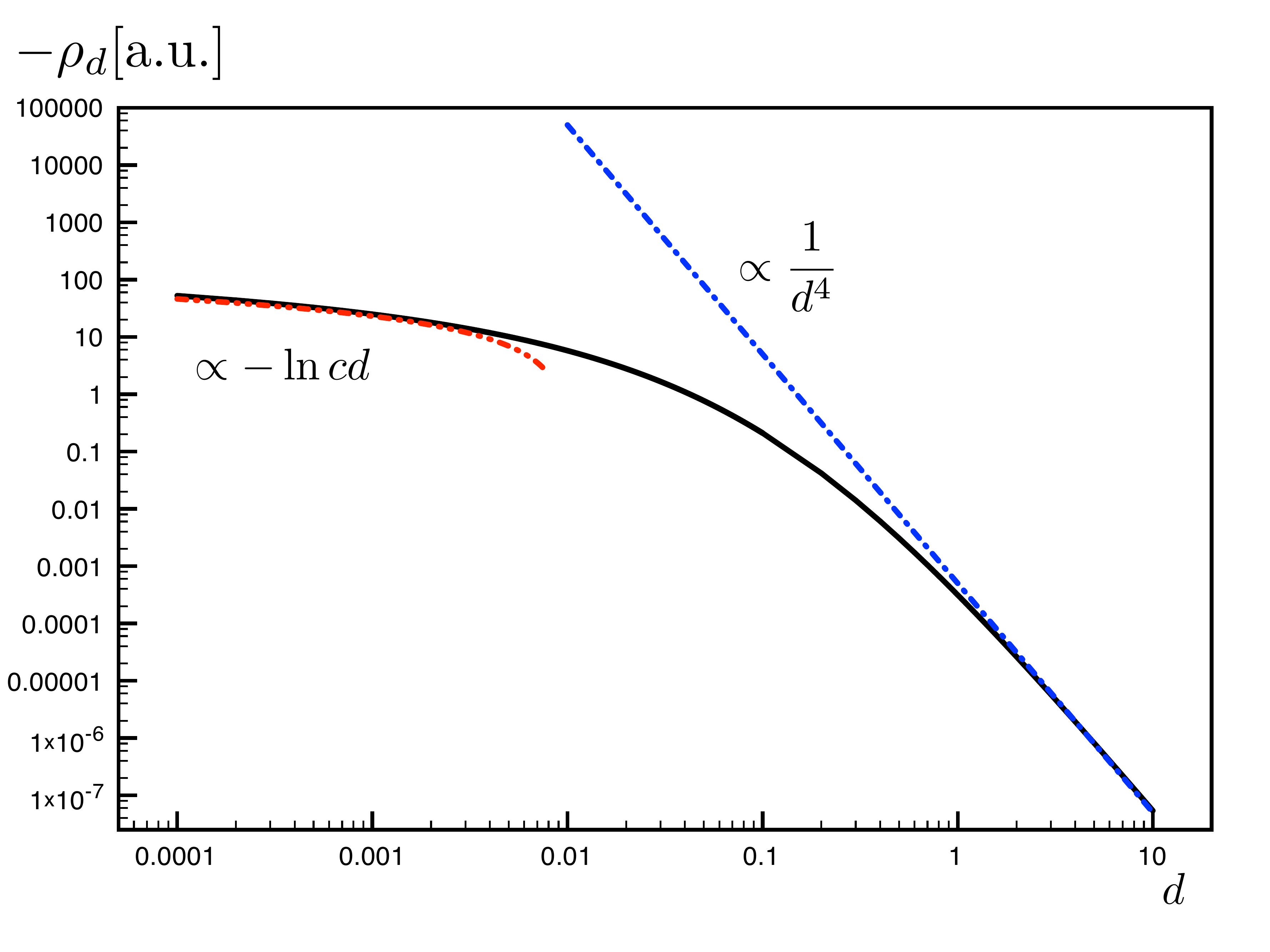}
\caption{Crossover from $\ln d$- to $1/d^4$-behavior in the $\overline{d}\ll 1$. The curve was obtained for $k_F=10$ and $T=0.01$. The crossover takes place roughly at $k_F d \approx 1$.}\label{fig:d}
\end{figure}

We have checked both statements against numerically integrating Eq.~\eqref{eq:rhodragfl} and the results are shown in Fig.~\ref{fig:d}.

\subsection{Drag in limit $\overline{d} \gg 1$}

This limit only allows for one limit of $\kappa$, since by construction the parameter which enters is given by $\overline{d} \overline{\mu}$, which by construction is large in this limit (remember $\mu/T \gg 1$). Consequently we have $\kappa \ll1$ and taking into account the fact that the integral over momentum $q$ is cut off on the scale one due to the exponential factors we find
\begin{eqnarray}
g(\overline{d} \gg 1,\kappa \ll 1) &\approx&  \int_0^\infty  \frac{dq q^3 e^{-2q}}{\left(1-e^{-2q}\right)^2} 4 \int_{-q/\overline{d}}^{q/\overline{d}} d\omega \nonumber \\ &=& \frac{\pi^4}{15 \overline{d}}\;.
\end{eqnarray}

\begin{figure}[h]
\includegraphics[width=0.49\textwidth]{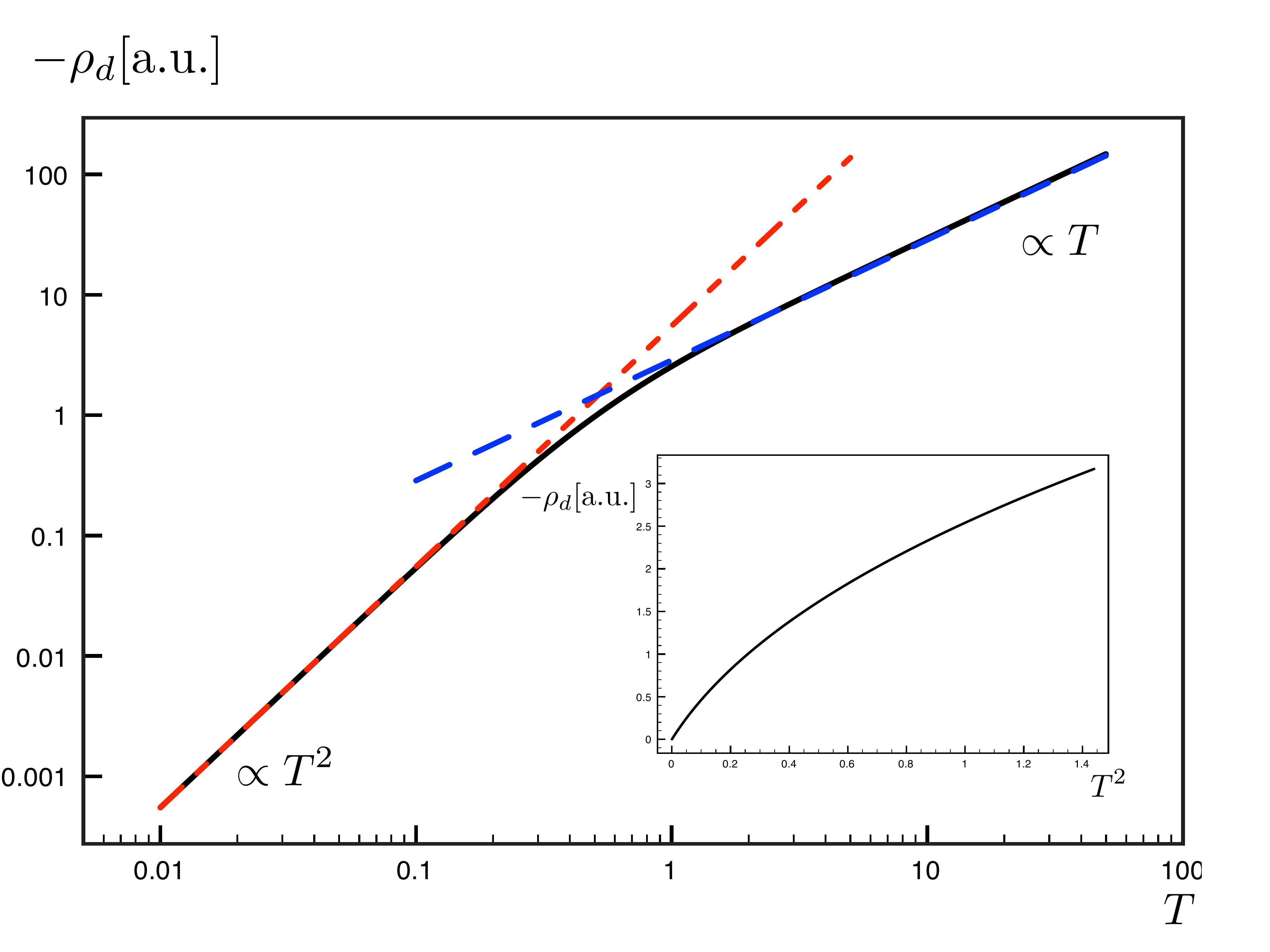}
\caption{Crossover from $T^2$ to $T$ linear behavior. The curve was obtained for $k_F=100$ and $d=1$. The crossover takes place roughly at $Td\approx 0.2$. This scale is achievable in currently available samples.}\label{fig:T}
\end{figure}

For the drag resistivity this implies
\begin{eqnarray}
\rho_d \approx \frac{h}{e^2} \frac{T}{\mu} \frac{16 \pi^5}{15 N^4} \frac{1}{\left(k_F d \right)^5 \alpha^2}\;,
\end{eqnarray}
meaning it is linearly proportional to $T$ and inversely proportional to $d^5$. Again, we have numerically integrated Eq.~\eqref{eq:rhodragfl} and verified these predictions, as shown in Fig.~\ref{fig:T}. We have found that the deviations from $T^2$-behavior become visible on the scale $\overline{d}\approx 0.1-0.2$. Most importantly, this shows that the use of Eq.~\eqref{eq:polarizationfl} as approximate form of the imaginary part of the polarization function is fully justified.

\subsection{Connection to experiments}

In the limit $\mu/T \gg 1$ we have identified a variety of regimes depending on whether $dT/v_F$ and $k_F d$ small or large
\begin{eqnarray}\label{eq:summary}
\rho_d \approx \left \{ \begin{array} {ccc}   \frac{h}{e^2} \left ( \frac{T}{\mu}\right)^2 \frac{8 \alpha^2}{3 \pi} \ln \frac{\alpha N k_F d}{2\pi}  & \overline{d} \ll 1  &  k_F d \ll 1 \\  \frac{h}{e^2} \left ( \frac{T}{\mu}\right)^2 \frac{16 \pi^3 \zeta(3)}{N^4} \frac{1}{\left( k_F d\right)^4 \alpha^2} &  \overline{d}\ll 1 & k_F d \gg 1 \\  \frac{h}{e^2} \frac{T}{\mu} \frac{16 \pi^5}{15 N^4} \frac{1}{\left(k_F d \right)^5 \alpha^2} & \overline{d} \gg 1 & k_F d \gg 1   \end{array}  \right. \;.
\end{eqnarray}

We have studied Eq.~\eqref{eq:drag} numerically, especially in the limit $k_F d \gg 1$, to determine the crossover. We find that in this limit the deviations from the $T^2$-behavior become sizeable already at $\overline{d} \approx 0.1-0.2$, leading to a broad crossover region. For a sample with interlayer distance $d\approx 10$ nm this translates to a crossover temperature on the order $T_{\rm{cross}}\approx 150K$, meaning that the deviation from the standard $T^2$ behavior should be observable in current samples. This is particularly interesting in light of the results in the recent experiment by Kim {\it et al.}~\cite{kim2011}. In this work a substantial deviation from the $T^2$ behavior was found in the temperature range above $150-200 K$. However, we note that Kim {\it et al.} extract their temperature behavior from the maximum of the drag while here we did so deep within the Fermi liquid regime. We have solved Eq.~\eqref{eq:draganalytical} as well as the full problem with all particle sorts and modes also in the regime of the maximum of drag and found results fully compatible with the above discussion. We thus conclude this section by stating that a possible explanation of the experimental finding of a deviation from the standard $T^2$ Fermi liquid behavior could be that the experiment enters the very broad crossover regime where the behavior crosses over to the linear in $T$ behavior. It is also interesting to note that very similar behavior was found in one of the earliest experiments on two-dimensional electron gases by Gramila {\it et al.}~\cite{gramila1991} where for higher temperatures large deviations from $T^2$ were observed. We have checked that in their work the crossover scale is also roughly given by $\overline{d}=0.2$. 
In a more recent experiment by Gorbatchev {\it et al.}~\cite{geim2012} the authors found $T^2$ behavior of the drag resistivity, consistent with the standard Fermi liquid predictions. However, instead of $1/d^4$ or $1/d^0$ the authors find a $1/d^2$ behavior. One possible explanation is that the measurement takes place for values of $k_F d \approx \mathcal{O}(1)$, where there is a crossover from $1/d^4$ to $1/d^0$ behavior, see Eq.~\eqref{eq:summary}. Coincidentally, the intermediate range might appear as $1/d^2$. An alternative explanation follows the paper by Kamenev {\it et al.}~\cite{kamenev1995} which discussed the Fermi liquid regime. In the diffusive limit it was found that there should be a behavior which is $T^2$, but of the $1/d^2$ type.

\section{Full crossovers}\label{sec:crossovers}

\begin{figure} [th]
\begin{center}
\includegraphics[width=0.45\textwidth]{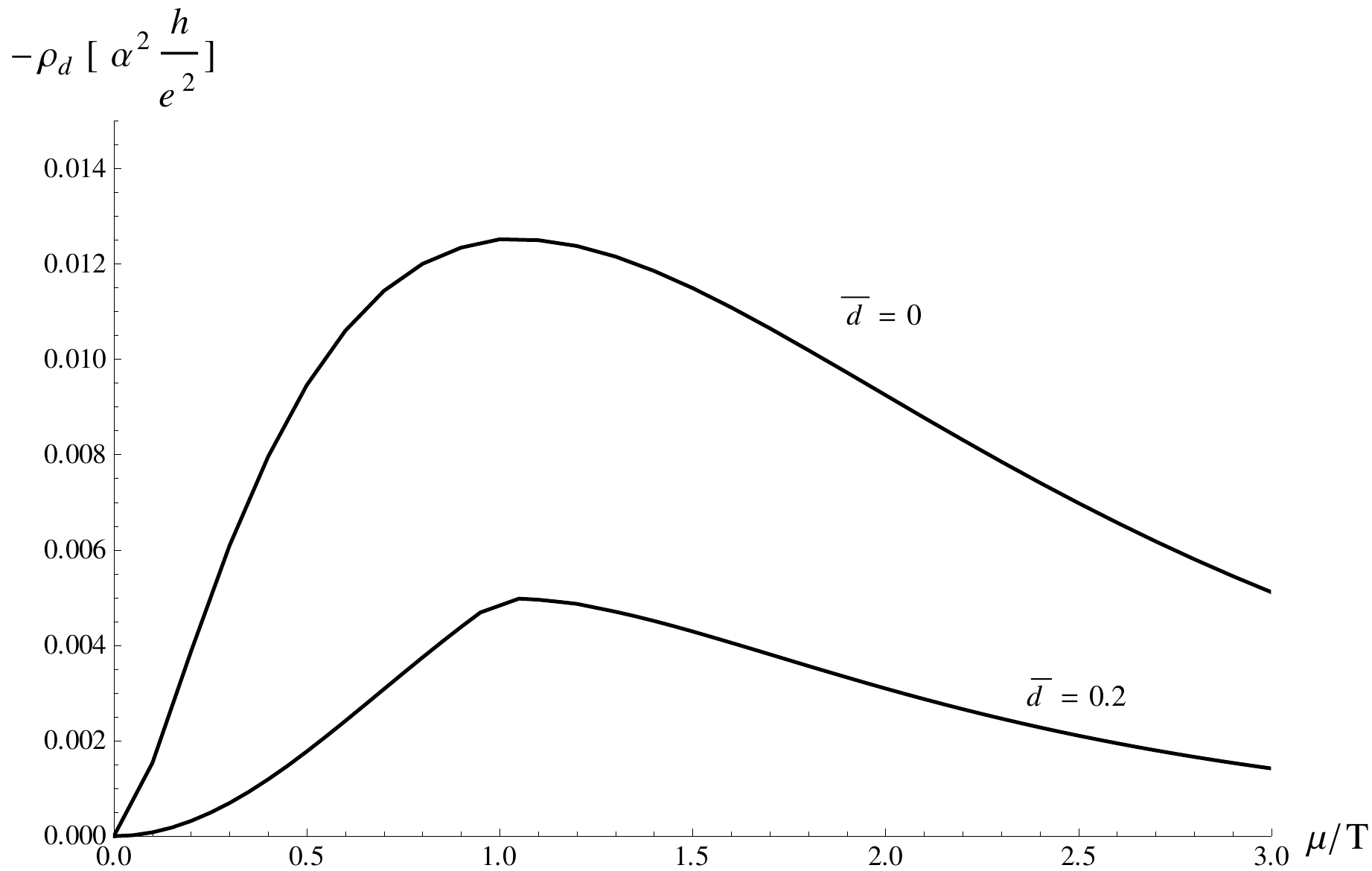}
\caption{Numerical evaluation of the drag resistivity for equal chemical potentials $ \mu_a =\mu_p = \mu $ and $\alpha =1 $.}\label{fig:crossover}
\end{center}
\end{figure}

So far we have discussed the limiting cases deep within the non-degenerate limit, $\mu/T \ll 1$, and in the opposite Fermi liquid regime where $\mu/T \gg1$. We conclude our discussions by commenting on the drag resistivity as it obtains in an experiment where both layers can be gated individually. This is relevant in experiments, where the charge density in the most general case varies in the two layers~\cite{kim2011,kim2012} even though more recent experiment achieve equal carrier densities with very high precision~\cite{geim2012}. This section mainly serves to complete the overall picture and for comparison to experimental findings. In this discussion we consider the fully screened interaction. In the case where both active and passive layers are kept at identical chemical potential sweeping the carrier density results in a curve which extrapolates between the $(\mu/T)^2$-behavior close to charge neutrality to the standard Fermi liquid regime, Fig.~\ref{fig:crossover}. For this plot we have kept both chemical potentials identical and show two different distances $\overline{d}$. We observe, as expected, a strong overall dependence of the order of magnitude as well as the position of the maximum with varying distance. 
Our numerical results and the overall picture are in good agreement with Refs.~\onlinecite{narozhny2011,polini2012}, which were obtained using a different formalism.

\begin{figure} [th]
\begin{center}
\includegraphics[width=0.45\textwidth]{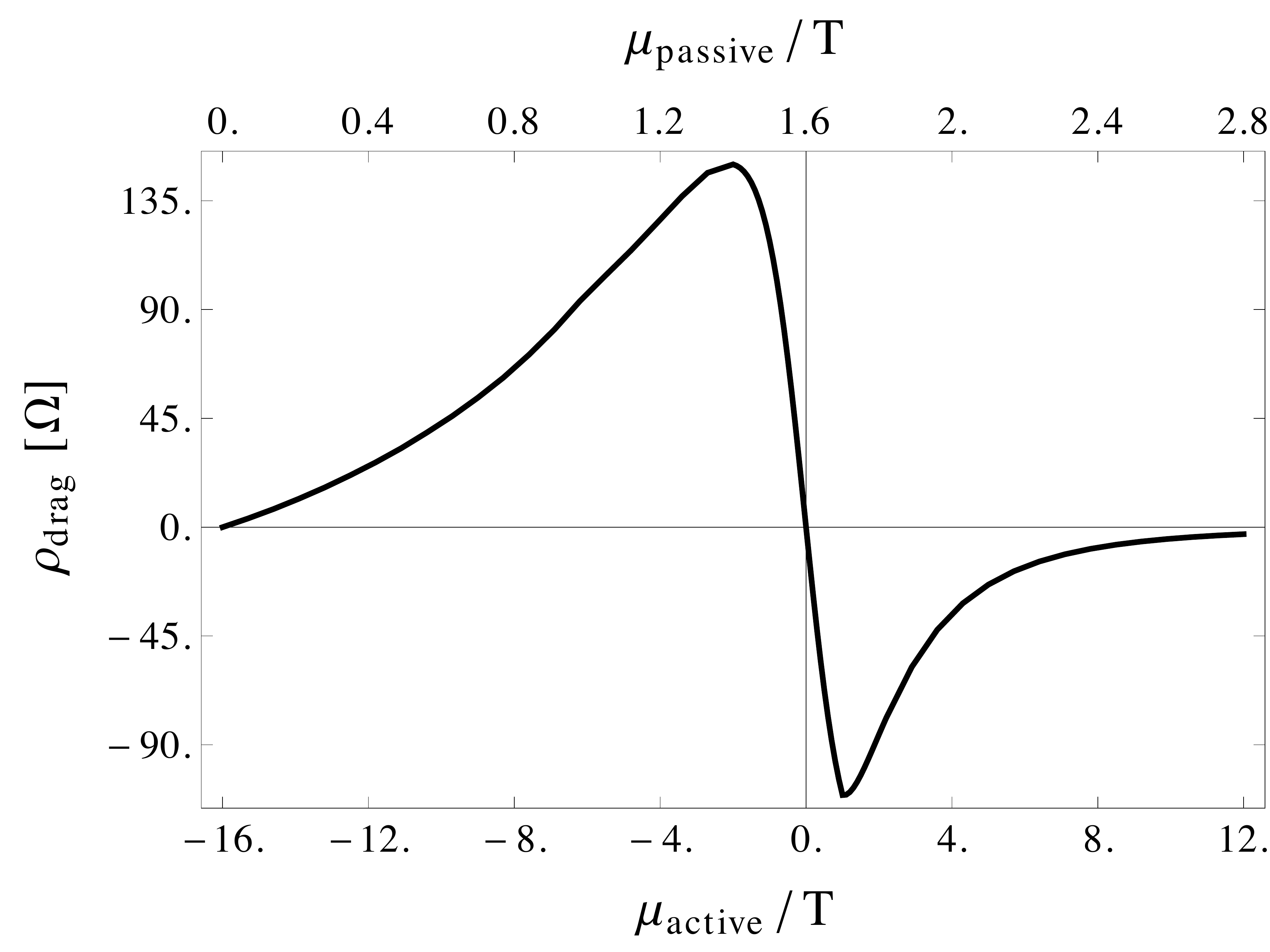}
\caption{Numerical modelling of the experiment by Kim {\it et al.}~\cite{kim2011} with $ \overline{d} =0.2 $ and $\alpha = 0.2$.} \label{fig:crossoverunequal}
\end{center}
\end{figure}

We have also analyzed the more realistic case in which both layers are not equally doped in Fig.~\ref{fig:crossoverunequal}. Here we have tried to emulate the experimental parameters of Kim {\it et al.}~\cite{kim2011, kim2012}.
We have choosen $ \overline{d} = 0.2 $ and $ \alpha =0.2 $ which seems to be fitting to their setup. For the dielectric constant of $ \text{Al}_2 \text{O}_3 $ we have chosen $\epsilon_r \approx 10$. The chemical potentials $\mu_a$ and 
$\mu_p $ are chosen such that we roughly reproduce the variation of densities as shown in Fig.~3 of Kim {\it et al.}~\cite{kim2011}. 
One can see that the numerical result is qualitatively in very good agreement with the experiment. In order to make quantitatively accurate comparison one would have to take into account the sample geometry, which we refrain from doing. 

\section{Conclusion}\label{sec:conclusion}

We have made an in depth study of Coulomb drag in two parallel monolayers of graphene in a variety of regimes ranging from the non-degenerate limit to the fully degenerate Fermi liquid limit. On a technical level we have employed a description of drag in terms of the Boltzmann kinetic approach using the variational approach within a two-mode approximation, which is asymptotically exact in the leading logarithmic approximation. We have studied the interplay between interactions, disorder, and the distance between the two monolayers. Directly at the Dirac point we find the absence of drag due to particle-hole symmetry. Still, we find that there is an interesting effect of the passive layer on the transport properties of the active layer, which comes from scattering of electrons and holes in the passive layer thereby relaxing a current. This effect is dependent on the parameter $Td/v_F$ and leads to an interesting temperature dependence of the inelastic scattering dominated single layer conductivity. We point out that this provides a promising route towards increasing inelastic scattering in graphene bringing closer the collision-dominated hydrodynamic limit. In the close vicinity of the Dirac point we found an interesting interplay between elastic and inelastic scattering. We first showed that in the clean limit at finite chemical potential there can be a well defined finite drag despite the divergence of all individual conductivities. We find that there is a non-commutativity of limits: first taking doping to zero and subsequently the disorder yields zero drag $\rho_d=0$ while in the reversed order of limits we find $\rho_d=-1/\sigma_0$, where $\sigma_0$ is the single layer conductivity of clean graphene at charge neutrality. This effect has been discussed by Sch\"utt {et al.}~\cite{schuett2012} in a related work of which we became aware during the completion of this manuscript. In the Fermi liquid regime, $\mu/T \gg1$, we presented a derivation of the drag resistivity in a very simplified setting of the Boltzmann equation which recovers the standard formula of drag as it has been derived in the context of Fermi liquids with the Kubo formalism. In the limit $T \to 0$ we find the standard Fermi liquid behavior of the $T^2$ type, with a distance dependence interpolating from $1/d^0$ for $k_F d\ll1$ to the more $1/d^4$-behavior in the opposite limit. Interestingly, we find a previously not discussed regime of $T$-linear behavior and $1/d^5$ distance dependence if $Td/v_F >1$. This behavior might be relevant for the understanding of recent experiments on graphene, where deviations from the $T^2$-behavior have been observed~\cite{kim2011,kim2012}. We point out that a similar behavior has also been seen in two dimensional electron gases and the crossover scales seem compatible~\cite{gramila1991}. We closed with a discussion of full crossover curves between both regimes which show good qualitative and quantitative agreement with experiments if screening is taken into account properly. 


\subsubsection*{Acknowledgements}
We acknowledge collaborations on related problems with M. M\"uller, S. Sachdev, and J. Schmalian as well as very useful discussions with M. Garst, A. Rosch, M. Sch\"utt, and M. Vojta. We are thankful to A. Geim and L. Ponomarenko for making unpublished material available to us.
This work was supported by the Emmy-Noether program FR 2627/3-1 (LF).

\appendix
\section{The scattering integral}\label{App:collision}

The interaction part of the scattering integral reads
\begin{widetext}
\begin{eqnarray}
&&I^{aa}_{\rm{C}}=2\pi\int \frac{d^{2}k_{1}}{
(2\pi )^{2}}\frac{d^{2}k_{2}}{(2\pi )^{2}}\Biggl\{  \notag \\
&&\delta (v_Fk-v_Fk_{1}-v_F|\mathbf{k}+\mathbf{q}|+v_F|\mathbf{k}_{1}-\mathbf{q}|)
R_{1}\Bigl\{f^a_{\lambda }(\mathbf{k},t)f^a_{-\lambda }(\mathbf{k}
_{1},t)[1-f^a_{\lambda }(\mathbf{k}+\mathbf{q},t)][1-f^a_{-\lambda }(\mathbf{k}_{1}-
\mathbf{q},t)]  \notag \\
&&~~~~~~~~~~~~~~~~~~~~~~~~~~~~~~~-[1-f^a_{\lambda }(\mathbf{k},t)][1-f^a_{-\lambda }(
\mathbf{k}_{1},t)]f^a_{\lambda }(\mathbf{k}+\mathbf{q},t)f^a_{-\lambda }(\mathbf{k}_{1}-
\mathbf{q},t)\Bigr\}  \notag \\
&&+\delta (v_Fk+v_Fk_{1}-v_F|\mathbf{k}+\mathbf{q}|-v_F|\mathbf{k}_{1}-\mathbf{q}|)
R_{2}\Bigl\{f^a_{\lambda }(\mathbf{k},t)f^a_{\lambda }(\mathbf{k}
_{1},t)[1-f^a_{\lambda }(\mathbf{k}+\mathbf{q},t)][1-f^a_{\lambda }(\mathbf{k}_{1}-
\mathbf{q},t)]  \notag \\
&&~~~~~~~~~~~~~~~~~~~~~~~~~~~~~~~-[1-f^a_{\lambda }(\mathbf{k},t)][1-f^a_{\lambda }(
\mathbf{k}_{1},t)]f^a_{\lambda }(\mathbf{k}+\mathbf{q},t)f^a_{\lambda }(\mathbf{k}_{1}-
\mathbf{q},t)\Bigr\}\Biggr\} \notag \\ &&I^{ap}_{\rm{C}}=2\pi \int \frac{d^{2}k_{1}}{
(2\pi )^{2}}\frac{d^{2}k_{2}}{(2\pi )^{2}}\Biggl\{  \notag \\
&&\delta (v_Fk-v_Fk_{1}-v_F|\mathbf{k}+\mathbf{q}|+v_F|\mathbf{k}_{1}-\mathbf{q}|)
\tilde{R}_{11}\Bigl\{f^a_{\lambda }(\mathbf{k},t)f^p_{-\lambda }(\mathbf{k}
_{1},t)[1-f^a_{\lambda }(\mathbf{k}+\mathbf{q},t)][1-f^p_{-\lambda }(\mathbf{k}_{1}-
\mathbf{q},t)]  \notag \\
&&~~~~~~~~~~~~~~~~~~~~~~~~~~~~~~~-[1-f^a_{\lambda }(\mathbf{k},t)][1-f^p_{-\lambda }(
\mathbf{k}_{1},t)]f^a_{\lambda }(\mathbf{k}+\mathbf{q},t)f^p_{-\lambda }(\mathbf{k}_{1}-
\mathbf{q},t)\Bigr\}  \notag \\
&&+\delta (v_Fk-v_Fk_{1}-v_F|\mathbf{k}+\mathbf{q}|+v_F|\mathbf{k}_{1}-\mathbf{q}|)
\tilde{R}_{12}\Bigl\{f^a_{\lambda }(\mathbf{k},t)f^p_{-\lambda }(\mathbf{k}
_{1},t)[1-f^p_{\lambda }(\mathbf{k}+\mathbf{q},t)][1-f^a_{-\lambda }(\mathbf{k}_{1}-
\mathbf{q},t)]  \notag \\
&&~~~~~~~~~~~~~~~~~~~~~~~~~~~~~~~-[1-f^a_{\lambda }(\mathbf{k},t)][1-f^p_{-\lambda }(
\mathbf{k}_{1},t)]f^p_{\lambda }(\mathbf{k}+\mathbf{q},t)f^a_{-\lambda }(\mathbf{k}_{1}-
\mathbf{q},t)\Bigr\}  \notag \\
&&+\delta (v_Fk+v_Fk_{1}-v_F|\mathbf{k}+\mathbf{q}|-v_F|\mathbf{k}_{1}-\mathbf{q}|)
\tilde{R}_{2}\Bigl\{f^a_{\lambda }(\mathbf{k},t)f^p_{\lambda }(\mathbf{k}
_{1},t)[1-f^a_{\lambda }(\mathbf{k}+\mathbf{q},t)][1-f^p_{\lambda }(\mathbf{k}_{1}-
\mathbf{q},t)]  \notag \\
&&~~~~~~~~~~~~~~~~~~~~~~~~~~~~~~~-[1-f^a_{\lambda }(\mathbf{k},t)][1-f^p_{\lambda }(
\mathbf{k}_{1},t)]f^a_{\lambda }(\mathbf{k}+\mathbf{q},t)f^p_{\lambda }(\mathbf{k}_{1}-
\mathbf{q},t)\Bigr\}\Biggr\} ,
\end{eqnarray}
where
\end{widetext}
\begin{eqnarray}
R_{1} &=&4N |T_{+--+}(\mathbf{k},\mathbf{k_{1}},\mathbf{q}
)|^{2}\notag \\ &&+|T_{+-+-}(\mathbf{k},\mathbf{k_{1}},\mathbf{k_{1}}-\mathbf{k}-\mathbf{
q})|^{2}   \notag \\
&&-4T_{+--+}(\mathbf{k},\mathbf{k_{1}},\mathbf{q})T_{+-+-}^{\star }(\mathbf{k
},\mathbf{k_{1}},\mathbf{k_{1}}-\mathbf{k}-\mathbf{q})  \notag \\
&&-4T_{+-+-}^{\star}(\mathbf{k},\mathbf{k_{1}},\mathbf{k_{1}}-\mathbf{k}-
\mathbf{q}
)T_{+--+}(\mathbf{k},\mathbf{k_{1}},\mathbf{q})\nonumber \\
\end{eqnarray}
and
\begin{eqnarray}
R_2&=&4N|T_{++++}(\mathbf{k},\mathbf{k_{1}},\mathbf{q})|^{2}
\notag \\
&&-4T_{++++}(\mathbf{k},\mathbf{k_{1}},\mathbf{q})T_{++++}^{\star }(\mathbf{k
},\mathbf{k_{1}},\mathbf{k_{1}}-\mathbf{k}-\mathbf{q})\;, \nonumber \\
\end{eqnarray}
while
\begin{eqnarray}
\tilde{R}_{11} &=&4N|\tilde{T}_{+--+}(\mathbf{k},\mathbf{k_{1}},\mathbf{q}
)|^{2} \nonumber \\ \tilde{R}_{12} &=&4N|\tilde{T}_{+-+-}(\mathbf{k},\mathbf{k_{1}},\mathbf{k_{1}}-\mathbf{k}-\mathbf{
q})|^{2} \nonumber \\ \tilde{R}_2&=&4N|\tilde{T}_{++++}(\mathbf{k},\mathbf{k_{1}},\mathbf{q})|^{2}\;.
\end{eqnarray}
We have used
\begin{eqnarray}
&&T_{\lambda_1 \lambda_2 \lambda_3 \lambda_4} (\mathbf{k}_1 , \mathbf{k}_2 ,
\mathbf{q}) = \frac{V({\bf q},\omega_{\mathbf{k}_1 ,\mathbf{q}})}{8} \times  \\ &\times& \left[ 1 +
\lambda_1 \lambda_4 \frac{(K_1^{\ast} + Q^{\ast}) K_1}{|\mathbf{k}_1 +
\mathbf{q}| k_1} \right] \left[1 + \lambda_2 \lambda_3 \frac{(K_2^{\ast} -
Q^{\ast}) K_2 }{|\mathbf{k}_2 - \mathbf{q}| k_2} \right] , \nonumber  \label{deft}
\end{eqnarray}
and
\begin{eqnarray}
&&\tilde{T}_{\lambda_1 \lambda_2 \lambda_3 \lambda_4} (\mathbf{k}_1 , \mathbf{k}_2 ,
\mathbf{q}) = \frac{U({\bf q},\omega_{\mathbf{k}_1 ,\mathbf{q}})}{8} \times  \\ &\times& \left[ 1 +
\lambda_1 \lambda_4 \frac{(K_1^{\ast} + Q^{\ast}) K_1}{|\mathbf{k}_1 +
\mathbf{q}| k_1} \right] \left[1 + \lambda_2 \lambda_3 \frac{(K_2^{\ast} -
Q^{\ast}) K_2 }{|\mathbf{k}_2 - \mathbf{q}| k_2} \right] . \nonumber  \label{inter}
\end{eqnarray}
In order to obtain $I^{pp}_{\rm{C}}$ and $I^{pa}_{\rm{C}}$ one simply has to change the individual indices. The collision integral due to disorder assumes the form
\begin{eqnarray}
I^{aa}_{\rm{dis}}&=&2\pi \int \frac{d^2 k_1}{(2\pi)^2} \delta(k-k_1) |U_{\lambda \lambda}({\bf{k}},{\bf{k}}_1)|^2 \times \nonumber \\  &\times& \Big[ f^a_\lambda (\vk,t)(1-f^a_\lambda(\vko,t))-(1-f^a_\lambda(\vk,t))f^a_\lambda (\vko,t) \Big] \nonumber \\
\end{eqnarray}
where again $I^{pp}_{\rm{dis}}$ is obtained from a simple change of indices.

\section{The scattering matrix}\label{App:scatt}
We define a space of modes according to

\begin{widetext}
\begin{eqnarray} \label{eq:def_modes}
\bg_{0 \l}^{a/p} (\bk)  =  \l \; \frac{\bk}{k} \; \chi^{a/p}_{0 \l} &
\text{ and } & \;
\bg_{1 \l}^{a/p} (\bk)  =  \frac{v_F}{T} \; \bk \; \chi^{a/p}_{1 \l} \; .
\end{eqnarray}
In this basis the elements of the scattering matrix assume the following forms
\begin{eqnarray}  \label{eq:scattering_matrix_aa}
C^{a a }_{i j, \l \l^\prime}
& = & \frac{2 \pi}{ T^2} \intk \int \frac{d^2 \bk_1}{(2 \pi)^2}
\int \frac{d^2 \bq}{(2 \pi)^2}  \; \delta (\l k + \l^\prime k_1 - \l
|\bk + \bq|- \l^\prime |\bk_1- \bq|) \times \nonumber \\
&& \times f^{0 a}_\l (k) f^{0 p}_{\l^\prime} (k_1) \left( 1-f^{0 a}_\l
(|\bk +\bq|) \right) \left( 1-f^{0 p}_{\l^\prime} (|\bk_1 - \bq|)
\right) \bg^{p}_{i \l} (\bk)  \times \nonumber \\
&& \times \Big[
R_1 \left( \delta_{\l \l^\prime} (\bg^{p}_{j \l^\prime} (\bk) -
\bg^{p}_{j \l^\prime} (\bk+\bq)) + (1-\delta_{\l \l^\prime}) (\bg^{p}_{j
\l^\prime} (\bq-\bk_1) - \bg^{p}_{j \l^\prime} (-\bk_1)) \right) +
\nonumber \\
&& + \delta_{\l \l^\prime} R_2 \left( \bg^{p}_{j \l^\prime} (\bk) -
\bg^{p}_{j \l^\prime} (\bk+\bq) +\bg^{p}_{j \l^\prime} (\bk_1) -
\bg^{p}_{j \l^\prime} (\bk_1-\bq) \right) +
\delta_{\l \l^\prime} \tilde{R}_{11}  \left( \bg^{p}_{j \l^\prime} (\bk)
- \bg^{p}_{j \l^\prime} (\bk+\bq) \right) + \nonumber \\
&&  + \tilde{R}_{12} \left( \delta_{\l \l^\prime} \bg^{p}_{j \l^\prime}
(\bk) - (1-\delta_{\l \l^\prime}) \bg^{p}_{j \l^\prime} (-\bk_1)  \right)
+ \delta_{\l \l^\prime} \tilde{R}_2 \left( \bg^{p}_{j \l^\prime} (\bk) -
\bg^{p}_{j \l^\prime} (\bk+\bq) \right) \Big] \; ,
\end{eqnarray}
and
\begin{eqnarray}  \label{eq:scattering_kernel_ap}
C^{a p }_{i j, \l \l^\prime}
& = & \frac{2 \pi}{ T^2} \intk \int \frac{d^2 \bk_1}{(2 \pi)^2}
\int \frac{d^2 \bq}{(2 \pi)^2}  \; \delta (\l k + \l^\prime k_1 - \l
|\bk + \bq|- \l^\prime |\bk_1- \bq|) \times \nonumber \\
&& \times f^{0 a}_\l (k) f^{0 p}_{\l^\prime} (k_1) ( 1-f^{0 a}_\l (|\bk
+\bq|) ) ( 1-f^{0 p}_{\l^\prime} (|\bk_1 - \bq|) ) \; \bg^{a}_{i \l}
(\bk)  \times \nonumber \\
&& \times \Big[ (1-\delta_{\l \l^\prime}) \tilde{R}_{11} \left(
\bg^{p}_{j \l^\prime} (\bk_1) - \bg^{p}_{j \l^\prime} (\bk_1-\bq) \right) +
\delta_{\l \l^\prime}  \tilde{R}_{2} \left( \bg^{p}_{j \l^\prime}
(\bk_1) - \bg^{p}_{j \l^\prime} (\bk_1-\bq) \right) + \nonumber \\
&& + \tilde{R}_{12} \left(  (1-\delta_{\l \l^\prime}) \bg^{a}_{j
\l^\prime} (\bq-\bk_1) -\delta_{\l \l^\prime} \bg^{p}_{j \l^\prime}
(\bk+\bq) \right)  \Big] \; .
\end{eqnarray}
$ C^{p p} $ and $ C^{p a} $ are obtained by a simple exchange of $ a $
and $ p $. The full equation can be cast in the form
\begin{eqnarray}  \label{eq:drag_matrix}
\left( \begin{array}{c} {\vec{ D}}_a \\ 0  \end{array} \right) =
\left ( \begin{array} {cc}  C^{aa} & C^{ap} \\ C^{pa}  &  C^{pp}
\end{array} \right)\cdot \left( \begin{array}{c} \vec{\chi}_a  \\
\vec{\chi}_p  \end{array} \right) & \Rightarrow &
\left( \begin{array}{c} \vec{\chi}_a  \\ \vec{\chi}_p
\end{array} \right) =
\left ( \begin{array} {cc}  C^{aa} & C^{ap} \\ C^{pa}  &  C^{pp}
\end{array} \right)^{-1} \cdot \left( \begin{array}{c}  {\vec {D}}_a \\ 0
\end{array} \right)\;.
\end{eqnarray}

\end{widetext}

\end{document}